\documentclass[9pt,shortpaper,twoside,web]{ieeecolor}

\usepackage{etoolbox}\newtoggle{IsTwocolumn}\settoggle{IsTwocolumn}{true}
\usepackage{generic_}

\usepackage{cite}
\usepackage{amsmath,amssymb,amsfonts}
\usepackage{algorithmic}
\usepackage{graphicx}
\usepackage{algorithm,algorithmic}

\usepackage{textcomp}

\usepackage{mathrsfs}   
\usepackage{color}
\usepackage{multirow}
\usepackage{booktabs} 
\usepackage{array} 

\usepackage{tikz}
\usetikzlibrary{arrows.meta,bending,shadows,shapes.callouts}
\usetikzlibrary{calc} 

\usepackage{hyperref}
\hypersetup{hidelinks=true}
                            
\newcommand{\delay}{{h}}
\newcommand{\Vvec}{{v}}

\newcommand{\out}{{\zeta}}

\newcommand{\rev}[1]{{#1}}

\allowdisplaybreaks 

\newtheorem{thm}{Theorem}[section]
\newtheorem{defn}[thm]{Definition}
\newtheorem{rem}[thm]{Remark} 
\newtheorem{exmp}[thm]{Example} 
\newtheorem{cor}[thm]{Corollary} 
\newtheorem{lem}[thm]{Lemma} 
\newtheorem{prop}[thm]{Proposition}

\def\BibTeX{{\rm B\kern-.05em{\sc i\kern-.025em b}\kern-.08em
    T\kern-.1667em\lower.7ex\hbox{E}\kern-.125emX}}

\makeatletter
    \def\@endtheorem{
		 \vspace{0pt}\hfill  $\blacktriangleleft$ 
		\endtrivlist\@endpefalse }
\makeatother

\iftoggle{IsTwocolumn}{
}{\setlength{\parskip}{2.5ex}}

\iftoggle{IsTwocolumn}{
}{
\usepackage{titlesec}
\titleformat{\section}
  {\large\scshape\sffamily\color{nblue}}{\thesection}{1em}{}
	}

\begin{document}
\title{\!Lyapunov--Krasovskii Functionals of 
Robust Type  \\
  for the Stability Analysis in Time-Delay Systems}
\author{Tessina H. Scholl
\thanks{The author is with the Institute for Automation and Applied Informatics, Karlsruhe Institute of Technology, 76021 Karlsruhe, Germany (e-mail:\ tessina.scholl@kit.edu). }
}

\maketitle

\begin{abstract}
Inspired by the widespread concept of Lyapunov--Krasovskii functionals of complete type, this article proposes an alternative class of functionals, termed Lyapunov--Krasovskii functionals of robust type. 
Their construction aims \rev{to improve resulting} robustness bounds of linear systems with a constant delay. These refer to bounds on nonlinear or uncertain terms that can be added to the system without compromising the proof of stability. \rev{The article derives, via the classical Lyapunov--Krasovskii theorem, an $H_\infty$-norm-based small-gain robustness result established by a functional of  known structure.}  
While complete-type functionals are related to infinite-dimensional Lyapunov equations, the proposed functionals are related to infinite-dimensional algebraic Riccati equations. 
Existence of \rev{the} functional relies on the operator-valued Kalman--Yakubovich--Popov lemma, which is \rev{made} applicable due to a splitting approach. In particular, for any asymptotically stable nominal system, there exists a Lyapunov--Krasovskii functional of robust type that \rev{yields} a nonzero bound on admissible perturbations. An example \rev{illustrates the reduction in conservatism compared with} complete-type functionals.
\end{abstract}


%\iftoggle{IsTwocolumn}{
%}{\vspace{1.5cm}Additional single-column version of the manuscript, requested by IEEE for the ease of reviewing.\newpage}

\section{Introduction}
\label{sec:introduction}

\IEEEPARstart{T}{his}  \rev{article addresses the Lyapunov-based stability analysis of time-delay systems in the presence of nonlinear or uncertain terms. 
In  systems without time delay, a small-gain robustness bound given by the  reciprocal of the $H_\infty$-norm  of the nominal linear system is well-known to be derivable via Lyapunov arguments. 
In particular, this robustness result comes with a   converse Lyapunov statement, i.e.,  the existence of a corresponding Lyapunov function is guaranteed, and it is known to be realizable in the form $V(x)=x^\top P x$ with some matrix $P$  \cite{Boyd.1994}.  
In contrast, for time-delay systems, to the best of our knowledge, no $H_\infty$-norm-based robustness result  is  available that 
guarantees the existence of a  structurally known functional satisfying the conditions of  the classical Lyapunov--Krasovskii (LK) theorem (which is, e.g., given in \cite[Thm.~5-1.2]{Hale.1993}).}  

\rev{So far, such a converse Lyapunov result  is only available for 
complete-type \cite{Kharitonov.2013,Kharitonov.2003} and related \cite{Medvedeva.2015}  LK functionals---which do not pertain to  $H_\infty$. Also recent existence guarantees for   LK functionals from semidefinite-programming-based stability criteria for time-delay systems \cite{Bajodek.2024} rely on proving that a complete-type or related functional can be recovered close enough once the number of free parameters is large enough. Note that} complete-type and related LK functionals establish  an active field of research \cite{Zhabko.2021,Juarez.2020b,Kudryakov.2023,Juarez.2020,Mondie.2022,Alexandrova.2020,Rychkov.2025}.

The question of robustness   is even the original purpose of complete-type LK functionals \cite{Kharitonov.2003}. 
The LK functional is constructed for  a nominal linear time-delay system  that is already known to have an asymptotically stable equilibrium. This nominal system  is described by 
\begin{align}\label{eq:unperturbed}
 \dot x(t)=A_0 x(t)+A_1 x(t-\delay)\quad  =: f(x_t)  ,   
\end{align} 
with $A_0,A_1\in \mathbb R^{n\times n}$, delay $\delay>0$, and  state  $x_t\in C([-\delay,0],\mathbb R^n)$, $x_t(\theta)=x(t+\theta), \forall t\geq 0$. Though constructed for (\ref{eq:unperturbed}), the LK functional is employed to prove stability in perturbed versions of~(\ref{eq:unperturbed}). 
Perturbations of interest are added  nonlinear or uncertain terms  in 
\begin{align} \label{eq:perturbed_RFDE}
\dot x(t) &= 
\underbrace{A_0 x(t)+ A_1 x(t-\delay)}_{f(x_t)}  +\underbrace{\tilde g\big(x(t),x(t-\delay)\big)}_{g(x_t)}
\end{align}
(relying on the same delay $h$). 
\rev{The  robustness statement to be obtained specifies} admissible perturbations ${\tilde g\big(x(t),x(t-\delay)\big)}$ 
that do not compromise the asymptotic stability previously observed for the zero equilibrium in~(\ref{eq:unperturbed}).

\rev{However, being related to an infinite-dimensional Lyapunov equation,} the robustness statements obtained from complete-type LK functionals in \cite{Kharitonov.2003,MelchorAguilar.2007,Kharitonov.2013} are  rather conservative. Despite improvements, the same holds for the approaches in \cite{Medvedeva.2015b,Alexandrova.2020,Villafuerte.2007}, which do not rely on complete-type but  on related LK functionals.  \rev{The gap to an $H_\infty$-norm-based bound can be significant.    
Notably, the concept of complete-type LK functionals~\cite{Kharitonov.2003} for robustness analysis has been introduced much later  than} the frequency-domain results from \cite{Popov.1962}, {\cite[Thm.\ 4.16' and 4.16'']{Halanay.1966}}, \cite[eq.~(7) et seq.]{Kato.1970b},  \cite{Walker.1967b,Bliman.2000b,Bliman.2002c}, \cite{Pritchard.1989}. \rev{The considerable interest and research effort in these  functionals, despite their significantly more conservative robustness results, underscore the strong demand for an LK-based approach. Altogether, to resolve this issue of  conservatism while retaining the advantages of LK functionals of complete type is the main objective of this article.}

\rev{To this end,} this article introduces an alternative class of quadratic LK functionals---termed LK functionals of robust type. 
Along the lines of LK functionals of complete type, LK functionals of robust type are  defined in terms of their derivative along solutions of the nominal system~(\ref{eq:unperturbed}). \rev{In contrast to an infinite-dimensional Lyapunov equation, the proposed concept is related to an infinite-dimensional algebraic Riccati equation. As a main result, a  small-gain robustness result is obtained that---via the classical LK theorem---yields the reciprocal of the $H_\infty$-norm of the nominal system as bound on the norm of admissible perturbations. An example at the end of this article demonstrates that the improvement compared to bounds obtained from complete-type LK functionals can be significant. The proposed LK functional of robust type  is guaranteed to exist,  has a known structure, and satisfies the classical LK conditions.}

\rev{A key idea in the presented derivation is the introduction of an  offset function  that induces a refinement of the constraint on $\tilde g$. This refinement is tailored to the specific requirements on the derivative of the LK functional, while the defining equation for the functional remains particularly  simple.}  

\rev{A key advantage of the proposed Lyapunov-based approach over pure frequency-domain small-gain arguments is that it can accommodate non-global perturbation constraints. 
That is, even  if the obtained robustness bounds are only met locally by $\tilde g$, a regional stability result can be  obtained if a structurally known functional is available.   
The resulting estimation of the domain of attraction is  beyond the scope of the present article, but the reader is referred to}   \cite{MelchorAguilar.2007,Villafuerte.2007,Alexandrova.2020}  for corresponding regional results from complete-type and related LK functionals.

\rev{A key challenge for obtaining an $H_\infty$-norm-based result via Lyapunov arguments is that, even if an embedding to the Hilbert space $L_2\times \mathbb R^n$ (which is not the considered state space\footnote{\rev{The  LK theorem  \cite{Hale.1993}  addresses the  state space of continuous functions~$C$. 
Classical results like linearizability or LaSalle's invariance principle are available for $C$, 
which is why  $C$ is preferred for nonlinear systems.}})
is used, the infinite-dimensional Kalman--Yakubovich--Popov (KYP) lemma from \cite{Likhtarnikov.1977} (see also \cite{Likhtarnikov.1977b,Likhtarnikov.1976}) is not applicable:}  To cope with the pointwise delayed term in $\tilde g$ from (\ref{eq:perturbed_RFDE}), an unbounded output operator would be required. Instead of weakening the lemma conditions, cf.~\cite{Anikushin.2023}, the present article overcomes that obstacle \rev{by  not using the KYP lemma for proving existence of the overall functional. Instead, the KYP lemma is only used for proving existence of a certain part of the functional.  
It is also  due to this splitting approach that  not only existence  but also the  structure of the functional is known.}

Results of this article were previously presented in the author’s doctoral dissertation \cite{Scholl.2024c}.  A numerical approach to the proposed concept was presented in~\cite{Scholl.2024b}.

\textit{Structure.}
The article is organized as follows. 
\rev{After discussing preliminaries in Sec.~\ref{sec:problemStatement}, Sec.~\ref{sec:LKFunctionalsOfRobustType} provides the definition of LK functionals of robust type. 
Sec.~\ref{sec:MonotonicityInPerturbedRFDE},  Sec.~\ref{sec:properties}, and Sec.~\ref{sec:PerturbationRestrictions}      establish the properties required by the LK theorem. A resulting robustness result  (small-gain) and an example are  discussed in Sec.~\ref{sec:smallgainlike} and \ref{sec:example}, before Sec.~\ref{sec:Conclusion} concludes the article.} 

\textit{Notation.}
Given $x\in \mathbb R^n$,  $\|x\|$ is an arbitrary norm on~$\mathbb R^n$,   $\|x\|_2$ the Euclidean norm. Further notations are the zero vector $0_n\in \mathbb R^n$, zero matrix $0_{n\times m}\in \mathbb R^{n\times m}$ (in short $0$), and identity matrix $I_n\in \mathbb R^{n\times n}$ (in short $I$).  
Positive \mbox{(semi-)}\allowbreak definiteness is addressed by $Q\succ 0_{n\times n}$ ($Q\succeq 0_{n\times n}$), implicitly requiring $Q=Q^H$. 
Square integrable $\mathbb C^n$-valued functions are $L_2([a,b],\mathbb C^n)$ or $L_2$. $M_2$ is the product space $M_2=L_2\times \mathbb C^n$, and $\mathscr A^*$ is the adjoint of an operator $\mathscr A$. $AC$ stands for absolutely continuous.  
Continuous $\mathbb R^n$-valued functions on   $[a,b]$ are denoted by $\phi\in C([a,b],\mathbb R^n)$ or $\phi\in C$  with $\|\phi\|_{C}=\max_{\theta\in [a,b]} \|\phi(\theta)\|$ and with  zero function  $0_{n_{[a,b]}}\!\in\! C([a,b],\mathbb R^n)$. The interior of a set $\Omega$ is   $\mathrm{int}(\Omega)$. 
See \cite[Sec.~5.2]{Hale.1993} for the formal definition of the derivative $D_{f}^+V:C\to\mathbb R$ of $V:C\to\mathbb R$ along   solutions of $\dot x(t)=f(x_t)$. 
The set of class-K functions is $\mathcal K=\{\kappa\in C([0,\infty),\mathbb R_{\geq 0}): {\kappa(0)=0},\text{ strictly increasing}\}$, and $\mathcal K_\infty=\mathcal K \cap \{\kappa: \kappa(s)\to \infty, s\to\infty\}$.

\section{Problem statement and preliminaries} \label{sec:problemStatement}

Consider the retarded functional differential equation (RFDE)  given in (\ref{eq:perturbed_RFDE}), which is decomposed into a linear part  $\dot x(t)=f(x_t)$ 
and a possibly nonlinear term 
$g(x_t)$, referred to as perturbation.  
\rev{According to (\ref{eq:perturbed_RFDE}), both may rely on $x(t)=x_t(0)$ and $x(t-\delay)=x_t(-\delay)$. The perturbation $g(x_t)$ might,  for instance,   \begin{itemize}
\item  be the remainder of a linearization, 
\item  involve a saturation term, or 
\item  address additive uncertainties in $A_0$ and $A_1$. 
\end{itemize}
Throughout the article it is assumed} that 
$g(0_{n_{[-\delay,0]}})=0_n$, which causes  $0_{n_{[-\delay,0]}}$  to be an equilibrium of (\ref{eq:perturbed_RFDE}). Moreover,  
$g$ is assumed to be
Lipschitz continuous on any bounded set, which is a sufficient property for well-posedness of (\ref{eq:perturbed_RFDE}),  see \cite{Hale.1993}. 

LK functionals from the present article are intended to be used with the classical LK theorem, recalled below. 
\begin{thm}[LK theorem {\cite[Thm.~5-1.2]{Hale.1993}}] \label{thm:LK}
If there exists a continuous   $V:C([-\delay,0],\mathbb R^n)\to \mathbb R$, upper bounded by  $\exists \kappa_2\in \mathcal K, \forall \phi\in C:V(\phi)\leq \kappa_2(\|\phi\|_C)$, and if there exists   
$\Omega \subseteq C([-\delay,0],\mathbb R^n)$ with $0_{n_{[-\delay,0]}}\in \mathrm{int}(\Omega)$ \rev{such that}
 \begin{align}\label{eq:partPosDef}
 \exists \kappa_1&\in \mathcal K, \forall \phi\in \Omega: &\quad  
\kappa_1(\|\phi(0)\|)&\leq V(\phi), 
\\ 
\label{eq:partNegDef}
 \exists \kappa_3&\in \mathcal K, \forall \phi\in \Omega: & \quad D_{(f+g)}^+V(\phi)&\leq -\kappa_3(\|\phi(0)\|),  
\end{align}
then the zero equilibrium of (\ref{eq:perturbed_RFDE}) 
 is locally asymptotically stable. Moreover, if $\Omega=C([-\delay,0],\mathbb R^n)$ and $\kappa_1\in \mathcal K_\infty$, then it is globally asymptotically stable. 
\end{thm}

In particular, the present article aims at quadratic LK functionals $ V\colon C([-\delay,0],\mathbb R^n)\to \mathbb R$ 
that 
have the form 
\begin{align}\label{eq:V}
V(\phi) 
&=
\phi^\top\!(0) \,P_{\mathrm{xx}} \,\phi(0)
+
2 \int_{-\delay}^0 \phi^\top\!(0)\, P_{\mathrm{xz}}(\eta) \hspace{0.5pt}\phi(\eta)\,\mathrm d \eta 
\iftoggle{IsTwocolumn}{ 
	\nonumber \\*
	&\quad +
}{\nonumber \\*	&\quad+}
\int_{-\delay}^0\int_{-\delay}^0 \phi^\top\!(\xi) P_{\mathrm{zz}}(\xi,\eta) \hspace{0.5pt}\phi(\eta)\,\mathrm d \eta\,\mathrm d\xi
\iftoggle{IsTwocolumn}{  
	\nonumber  \\*
	&\quad+
}{+} 
\int_{-\delay}^0 \phi^\top\!(\eta) P_{\mathrm{zz,diag}}(\eta) \hspace{0.5pt}
\phi(\eta)\,\mathrm d \eta ,
\end{align}
$P_{\mathrm{xx}}\in \mathbb R^{n\times n}$;  $P_{\mathrm{zz,diag}},P_{\mathrm{xz}}\in L_2([-\delay,0],\mathbb R^{n\times n})$; $P_{\mathrm{zz}}\in L_2([-\delay,0]\times [-\delay,0] ,\mathbb R^{n\times n})$. 

\rev{To make our objectives} more apparent, 
let us  briefly revisit 
 the key aspects of the existing concept of complete-type 
 LK functionals. 
These functionals are defined 
in terms of 
$D_f^+V(x_t)$, which is the derivative of $V$ along solutions of   the nominal linear part 
(\ref{eq:unperturbed})
of (\ref{eq:perturbed_RFDE}). 
\begin{defn}[Complete-type LK functional  {\cite[Def.~2.7]{Kharitonov.2013}}] 
\label{def:complete-type}
A functional $V\colon C([-\delay,0],\mathbb R^n)\to \mathbb R$ 
that has the structure~(\ref{eq:V}) is called an 
  {\itshape LK functional of complete type} w.r.t.\ 
\begin{itemize}
\item the nominal linear system $\dot x(t)=f(x_t)$ and  
\item chosen real matrices $Q_0,Q_1,Q_2\succ 0_{n\times n}$
\end{itemize}
  if, for all 
  $ \phi\in C([-\delay,0],\mathbb R^n)$,  it holds 
\begin{align} \label{eq:determining_eq_DfV_completeTypeLK}
D_f^+ V(\phi) 
&= -\phi^\top\! (0)\, Q_0\, \phi(0) - \phi^\top\!(-\delay)\,Q_1\,\phi(-\delay)
\iftoggle{IsTwocolumn}{
	\nonumber \\*
	&\quad 
	-
}{-}
\int_{-\delay}^0 \phi^\top (\theta)\, Q_2 \,\phi(\theta)\,\mathrm d \theta. 
\\[-2.5em]\nonumber
\end{align}
\end{defn}
 
\rev{Note that (\ref{eq:determining_eq_DfV_completeTypeLK}) prescribes $D_f^+V$. However, decisive for (\ref{eq:perturbed_RFDE}) is whether   $D_{(f+g)}^+V$ still admits an upper bound (\ref{eq:partNegDef}). 
The latter can be confirmed if} the perturbation $g$ in (\ref{eq:perturbed_RFDE}) is small enough in the sense of\footnote{Note that 
the bound (\ref{eq:completeLK_linearNormBound}) 
resembles the well-known Lyapunov-equation-based perturbation bound for  delay-free systems $\dot x=A_0x+g(x)$ with $A_0\in \mathbb R^{n\times n}$  Hurwitz, stating that a perturbation $\|g(x)\|\leq \tilde \gamma  \|x\|$ is admissible if $\tilde \gamma<   \frac{\lambda_{\min}(Q_0)}{2 \lambda_{\max}(P)}$ from 
$PA_0+A_0^\top P=-Q_0$, see \cite[p.~140]{Khalil.2002}. } 
\begin{align}\label{eq:completeLK_linearNormBound}
\|g(\phi)\|_2
&\leq \tilde \gamma \left\|\left[\begin{smallmatrix} \phi(-\delay) \\ \phi(0)\end{smallmatrix}\right]\right\|_2 
\\*
 \text{with }
\tilde \gamma
&<
\min\big\{ 
\tfrac {\lambda_{\min} (Q_{0})}{2+\delay\|A_1\|_2}, 
\tfrac {\lambda_{\min} (Q_{1})}{1+\delay\|A_1\|_2},
\tfrac {\lambda_{\min} (Q_{2})}{\|A_1\|_2}
\big\}
\big/
\lambda_{\max}(\Psi(0))
,  
\nonumber
\end{align}
see \cite[Thm.~1]{MelchorAguilar.2007}, and also \cite[Thm.~6]{Kharitonov.2003}, \cite[Thm.~2.17]{Kharitonov.2013}, where $\Psi\colon \mathbb R\to \mathbb R^{n\times n}$ is the delay Lyapunov matrix function of (\ref{eq:unperturbed}) associated with $Q_0+Q_1+\delay Q_2$, see \cite[Def.~2.5]{Kharitonov.2013}.

\begin{thm}[\rev{Robustness statement, complete-type}]
\label{thm:robustness}
Let the zero equilibrium in (\ref{eq:unperturbed}) be asymptotically stable. Then the zero equilibrium of the perturbed system (\ref{eq:perturbed_RFDE}) is still globally asymptotically stable if  the perturbation $g$ in (\ref{eq:perturbed_RFDE}) is such that it satisfies the linear norm bound (\ref{eq:completeLK_linearNormBound})   for any $\phi\in C$. In particular, there exists a corresponding LK functional $V$ that satisfies the conditions from the  classical LK theorem (Thm.~\ref{thm:LK})  and that has the form (\ref{eq:V}). Such a functional $V$ is always found by  solving~(\ref{eq:determining_eq_DfV_completeTypeLK}). 
\end{thm}
\begin{proof}
\rev{ All conditions of Thm.~\ref{thm:LK} are met: Existence of a solution  of (\ref{eq:determining_eq_DfV_completeTypeLK})  \cite[Thm.~2.8, 2.11]{Kharitonov.2013};  Lower bound on $V$  \cite[Lem.~2.10]{Kharitonov.2013}; Upper bound on $D_{(f+g)}^+V$ if   (\ref{eq:completeLK_linearNormBound}) holds  \cite[Thm.~1]{MelchorAguilar.2007}.} 
\end{proof}
The present article introduces an alternative class of LK functionals. Similar to Def.~\ref{def:complete-type}, the defining equation shall be expressed in terms of $D_f^+V$. As $D_f^+V$ will have a different structure than (\ref{eq:determining_eq_DfV_completeTypeLK}), the proposed functionals 
 do not belong to 
the well-known class of complete-type LK functionals. 
Still, similar to above, \rev{the requirements of the LK theorem shall be met} whenever the equilibrium of the nominal linear system is asymptotically stable.  \rev{We aim to achieve the latter under a linear norm bound on $g$ that is less restrictive than (\ref{eq:completeLK_linearNormBound}).}

\section{Definition of LK functionals of robust type}\label{sec:LKFunctionalsOfRobustType}

\rev{To  enable  bounds that can be  tailored to  a specific structure of the perturbation $g(\phi)=\tilde g(\phi(0),\phi(-\delay))$ in (\ref{eq:perturbed_RFDE}), three matrices $B$, $C_0$, and $C_1$ are to be chosen. Let 
\begin{align}\label{eq:C_def}
 B\in \mathbb R^{n\times m}, C_0\in \mathbb R^{p_0\times n}, C_1\in \mathbb R^{p_1\times n},   \mathcal C \phi :=  
\left[\begin{matrix}
 C_1  \phi(-\delay) \\ C_0 \,\phi(0)
\end{matrix}\right], 
\end{align} 
and $p_0+p_1=:p$. If, as in (\ref{eq:completeLK_linearNormBound}),  no   information about the perturbation structure shall be taken into account, we choose  
\begin{align}\label{eq:unstructured}
B=I_n^{}, \qquad C_0=I_n^{}, \qquad  C_1=I_n^{}. 
\end{align}
Otherwise, the  matrices are intended to mark that   a bound on the specifically acting core perturbation function $a\colon \mathbb R^p\to \mathbb R^m$} from\footnote{
The negative sign in (\ref{eq:g_decomp}) is intended to resemble a negative feedback, being a common representation of 
Lur'e systems \cite{Khalil.2002}.  
}   
 \begin{align} \label{eq:g_decomp}
g(\phi)&= - B^{} \, a (\mathcal C \phi), 
\\
\rev{\text{or, equivalently,} \quad \tilde g\big(\phi(0),\phi(-\delay)\big)} &=  - B^{} \, a \left(\left[\begin{smallmatrix} C_1  \phi(-\delay) \\ C_0 \,\phi(0) \end{smallmatrix}\right]\right)
\nonumber 
\end{align}   
is of interest. \rev{Note that $a(0_p)=0_m$ since $\tilde g(0_n,0_n)=0_n$ by the assumption from Sec.~\ref{sec:problemStatement}.  A decomposition (\ref{eq:g_decomp}) is always possible in multiple ways. It can contribute to a more specific and thus less conservative robustness bound. We will particularly focus on full-rank choices for $C_0$ and $C_1$, see the  following example or~(\ref{eq:unstructured}).}

\begin{exmp}[Perturbation structure, choice]\label{ex:perturbationStructure}
\rev{Consider as an example $
g(x_t)=[0,x_1^2(t-\delay)]^\top$. Since $f(x_t)$ in (\ref{eq:perturbed_RFDE}) might be sensitive to perturbations in the first component,  the information that  the first component is not perturbed can be decisive. Thus, we use $B=[0,1]^\top$. If full-rank matrices $C_0,C_1$ are desired,  
\begin{align*}
g(x_t)= -\begin{bmatrix} 0 \\ 1\end{bmatrix} a(\mathcal C x_t ), 
\quad
\mathcal C x_t=\left[\begin{smallmatrix} x_1(t-\delay) \\ \varepsilon x_2(t-\delay) \\ \varepsilon x_1(t) \\ \varepsilon x_2(t) \end{smallmatrix}\right], \quad 
a(\zeta)=\zeta_1^2,
\end{align*}
i.e., 
$C_1=\big[\begin{smallmatrix}  1 & 0 \\ 0 & \varepsilon \end{smallmatrix}\big]$, $C_0=\varepsilon I_2$,  $\varepsilon\neq 0$, can  for instance be chosen.} 
\end{exmp}

\rev{Similarly to (\ref{eq:completeLK_linearNormBound}), in this article, we focus on  a linear norm bound  
\begin{align}\label{eq:linearNormBound}
\|a(\mathcal C \phi)\|_2\leq \gamma \|\mathcal C \phi\|_2,   
\end{align} 
on the core perturbation $a$,  with the admissible range of $\gamma>0$ being  of interest. Note that, if (\ref{eq:unstructured}) is chosen in (\ref{eq:g_decomp}) and thus  $a((\cdot,\cdot))=-\tilde g(\cdot,\cdot)$, then (\ref{eq:linearNormBound})  becomes a general bound on $g$ like (\ref{eq:completeLK_linearNormBound}).} 

Finally, the proposed defining equation for $D_f^+V$   depends  implicitly  on the solution $V$ itself, as,  
based on the first two terms 
from~(\ref{eq:V}),  
\begin{align} \label{eq:Vvec_perturbation}
v(\phi)&:=  P_{\mathrm{xx}} \,\phi(0) + \int_{-\delay}^0 P_{\mathrm{xz}}(\eta) \phi(\eta)\,\mathrm d \eta 
\end{align}
is encountered in the following definition.

\begin{defn}[LK functional of robust type] \label{def:LK_functional_of_robust_type}
A functional 
 $V\colon C([-\delay,0],\mathbb R^n)\to \mathbb R$ 
that has the structure (\ref{eq:V}) is called an 
{\itshape LK functional of robust type} w.r.t.\ \vspace{-0.25em}
\begin{itemize}
\item the nominal linear system $\dot x(t)=f(x_t)$, 
\item the perturbation structure $(B,\mathcal C)$, 
and
\item \rev{the  linear-norm-bound-type\footnote{\rev{Def.~\ref{def:LK_functional_of_robust_type} can be extended to other sector-type constraints  \cite[Def.~5.2.4]{Scholl.2024c}.}} 
constraint (\ref{eq:linearNormBound})} 
\end{itemize} 
\rev{if for all 
  $ \phi\in C([-\delay,0],\mathbb R^n)$  it holds 
\begin{align} \label{eq:determining_eq_DfV_linear_norm_bound}
D_f^+ V(\phi) 
= -\gamma^2 (\mathcal C \phi)^\top   \mathcal C\phi
  - \Vvec^\top\!(\phi) B 
  B^\top \Vvec(\phi)   
	\end{align}
with $\Vvec\colon C([-\delay,0],\mathbb R^n)\to \mathbb R^n$  given by  (\ref{eq:Vvec_perturbation}).}
\end{defn}
\rev{Notably,} 
the first term on the right-hand side of (\ref{eq:determining_eq_DfV_linear_norm_bound})  
\begin{align*}
-(\mathcal C \phi)^\top   \mathcal C \phi\stackrel{(\ref{eq:C_def})}= -\phi^\top\!(0) C_0^\top C_0 \phi(0) - \phi^\top\!(-\delay) C_1^\top C_1 \phi(-\delay) 
\end{align*}  
resembles  $-\phi^\top\!(0) Q_{0} \phi(0) - \phi^\top\!(-\delay) Q_{1} \phi(-\delay)$ from (\ref{eq:determining_eq_DfV_completeTypeLK}).
\rev{A numerical approach to solve (\ref{eq:determining_eq_DfV_linear_norm_bound})  is presented in \cite{Scholl.2024b}. However,  all properties of $V$ that are needed in this article} can be  accomplished without resorting to an explicit expression of~(\ref{eq:V}).  
\section{Domain with an upper bound on $D_{(f+g)}^+V$}
\label{sec:MonotonicityInPerturbedRFDE}

\rev{We start with the most important question---the counterpart to  (\ref{eq:completeLK_linearNormBound}): under which constraints on the perturbation in (\ref{eq:perturbed_RFDE}) can condition (\ref{eq:partNegDef}) on $D_{(f+g)}^+V$ be guaranteed if $D_f^+V$ is known to satisfy (\ref{eq:determining_eq_DfV_linear_norm_bound}).} 

The following intermediate result is \rev{independent from the defining equation (\ref{eq:determining_eq_DfV_linear_norm_bound}) but is a mere consequence of the structure (\ref{eq:V}) of $V$.} 

\begin{lem}[Perturbation effect on the derivative] \label{lem:D_perturbed}
Along  solutions of    (\ref{eq:perturbed_RFDE}), any functional of the form (\ref{eq:V}) has  the derivative 
\begin{align} \label{eq:D_fg_V}
D_{(f+g)}^+ V(\phi)
&=
D_f^+ V(\phi) + 2\, 
\Vvec^\top\!(\phi) \,g(\phi)
\end{align}
with $v$ being defined in (\ref{eq:Vvec_perturbation}).
\end{lem}
\begin{proof} 
For the special case of complete-type functionals, a detailed derivation is found in \cite[Lem.~2.14]{Kharitonov.2013}.  
The following consideration 
shows that (\ref{eq:D_fg_V})  holds for all functionals 
with the general 
structure (\ref{eq:V}). 
For $\phi=x_t$,   (\ref{eq:V}) becomes 
\begin{align}
V(x_t) 
&=
x^\top\!(t) P_{\mathrm{xx}} \,x(t)
+
2 x^\top\!(t) \smallint_{t-\delay}^t  P_{\mathrm{xz}}(\eta-t) \hspace{0.5pt}x(\eta)\,\mathrm d \eta
\nonumber \\* 
&\quad +
\smallint_{t-\delay}^t\smallint_{t-\delay}^t x^\top\!(\xi) P_{\mathrm{zz}}(\xi-t,\eta-t) \hspace{0.5pt}x(\eta)\,\mathrm d \eta\,\mathrm d\xi
\iftoggle{IsTwocolumn}{ 
	\nonumber  \\* 
	&\quad+ 
	}{+}
\smallint_{t-\delay}^t x^\top\!(\eta) P_{\mathrm{zz,diag}}(\eta-t) \hspace{0.5pt}x(\eta)\,\mathrm d \eta. \label{eq:V_xt}
\end{align}
 We compare $D_f^+V(x_t)$ and $D_{(f+g)}^+V(x_t)$, i.e.,   the derivative of (\ref{eq:V_xt}) 
along trajectories of  
$\dot x (t)= f(x_t)+\beta g(x_t)$ with   $\beta=0$ versus $\beta=1$.  
Writing out
\begin{align}
&D_{(f+\beta g)}^+ V(x_t)
=
2 \underbrace{\dot x^\top\!(t)}_{(f(x_t)+\beta g(x_t))^\top } P_{\mathrm{xx}} \,x(t)
\nonumber
\iftoggle{IsTwocolumn}{ 
	\\ 
	&+
}{+}
2 \Big(\underbrace{\dot x^\top\!(t)}_{\hspace{-2cm}(f(x_t)+\beta g(x_t))^\top\hspace{-2cm}}  \smallint_{t-\delay}^t  P_{\mathrm{xz}}(\eta-t) \hspace{0.5pt}x(\eta)\,\mathrm d \eta
+
 x^\top\!(t) \tfrac{\mathrm d}{\mathrm d t}
\smallint_{t-\delay}^t  
(\ldots) 
\,\mathrm d \eta
\Big)
\iftoggle{IsTwocolumn}{ 
	\nonumber\\ 
	&+
	}{	\nonumber\\	&\hspace{2.75cm} +}
\tfrac{\mathrm d}{\mathrm d t} \smallint_{t-\delay}^t\smallint_{t-\delay}^t (\ldots) \,\mathrm d \eta\,\mathrm d\xi
+ 
\tfrac{\mathrm d}{\mathrm d t}  \smallint_{t-\delay}^t (\ldots) \,\mathrm d \eta,  
\end{align}
the abbreviated terms  $\int_{t-\delay}^t (\ldots) \,\mathrm d \eta$ describe integrals from (\ref{eq:V_xt}) that, when differentiating via the Leibniz integral rule, do not give rise to $\dot x(t)$. Thus, they do not give rise to  $\beta$. 
Therefore, the scalar difference between  $D_f^+V(x_t)$ and $D_{(f+g)}^+V(x_t)$ 
 is obtained by 
$
2 g^\top\! (x_t) \Big(P_{\mathrm{xx}} \,x(t) + 
\smallint_{t-\delay}^t  P_{\mathrm{xz}}(\eta-t)  \hspace{0.5pt}x(\eta)\,\mathrm d \eta\Big)=2 g^\top\! (x_t) v(x_t)$. 
\end{proof}

\rev{Note that the aimed type of constraint (\ref{eq:linearNormBound}) on the core perturbation function $a$ is equivalent to} $a^\top\!(\mathcal C \phi)\, a(\mathcal C \phi)\leq \gamma^2 \;(\mathcal C \phi)^\top (\mathcal C  \phi)$ or 
\begin{align}
w\big(\,\mathcal C  \phi, \,a(\mathcal C  \phi) \,\big)&\geq 0  
\quad 
 \text{ with }\; 
w(\out,\alpha):=\gamma^2 \out^\top \out - \alpha^\top\alpha. \label{eq:linearNormBound_w}
\end{align}
\rev{The idea behind the defining equation 
(\ref{eq:determining_eq_DfV_linear_norm_bound}) 
is 
that} the  derivative $D_{(f+g)}^+V$ from (\ref{eq:D_fg_V}) 
can easily be shown to be nonpositive  for all  $\phi$  
for which $\mathcal C \phi\in \{\zeta\in \mathbb R^p: w(\zeta,a(\zeta))\geq 0\}$ from (\ref{eq:linearNormBound_w}). \rev{The latter is   realized choosing $\ell(\zeta)=0$ in the further-reaching theorem below.} 
\begin{thm}
[Upper bound on $D_{(f+g)}^+V$]
 \label{thm:monotonicity} 
Let $g$ in  (\ref{eq:perturbed_RFDE}) be decomposed according to  (\ref{eq:g_decomp}). 
Let $V$ be a functional of the form (\ref{eq:V}) that solves (\ref{eq:determining_eq_DfV_linear_norm_bound}).  
For any desired offset function  $\ell\colon \mathbb R^p\to \mathbb R$,  
 the derivative of $V$ along solutions of  
(\ref{eq:perturbed_RFDE}) satisfies 
\begin{align}\label{eq:kappa_3_existence}
 \forall \phi\in \Omega_{\ell}: \quad D_{(f+g)}^+ V (\phi)\leq - \ell(\mathcal C \phi ) 
\end{align}   
on $\Omega_{\ell}\subseteq C$ being 
the set  of all $\phi\in C$ 
for which the perturbation restriction (\ref{eq:linearNormBound_w})   
is exceeded  by   $\ell$    in the sense of 
\begin{align}
w(\mathcal C \phi ,a(\mathcal C \phi)) \geq \ell(\mathcal C \phi ). 
 \label{eq:perturbation_restriction_offset}
\\*[-2.25em]\nonumber
\end{align}
\end{thm}
\begin{proof}
Consider 
(\ref{eq:D_fg_V}) with $g(\phi)$ from (\ref{eq:g_decomp}), i.e., 
\begin{align}\label{eq:D_fBaC_V}
D_{(f+g)}^+V(\phi)
&=
D_f^+V(\phi) - 2v^\top\!(\phi) B a(\mathcal C \phi).
\end{align}
\rev{Abbreviating $\hat b^\top := v^\top\!(\phi) B$ in (\ref{eq:determining_eq_DfV_linear_norm_bound}), and $\hat a:=   a(\mathcal C \phi)$ in (\ref{eq:D_fBaC_V})   
yields 
\begin{align} 
D_{(f+g)}^+V(\phi)
&\stackrel{(\ref{eq:determining_eq_DfV_linear_norm_bound})}=
-\gamma^2 (\mathcal C \phi)^\top   (\mathcal C \phi) -\hat b ^\top \hat b 
 - 2v^\top\!(\phi) B  \hat a 
\nonumber 
 \\
&\stackrel{\hphantom{(00)}}=
-\gamma^2 (\mathcal C \phi)^\top   (\mathcal C \phi)    -\|\hat b + \hat a\|_2^2 
+
\hat a^\top \hat a  \nonumber  
\\
&\stackrel{(\ref{eq:linearNormBound_w})}=-w(\mathcal C \phi ,a(\mathcal C \phi) ) -\|\hat b + \hat a\|_2^2 .   \label{eq:DfgV_last_step}
\end{align}}Using (\ref{eq:perturbation_restriction_offset}), the result (\ref{eq:DfgV_last_step}) leads to the estimate (\ref{eq:kappa_3_existence}).  
\end{proof}
\rev{Let $\mathrm{rank}(C_0)=n$. Then, choosing a  class-K function 
$\ell(\zeta)=\kappa(\|\zeta\|)$, $\kappa\in \mathcal K$,  
in (\ref{eq:perturbation_restriction_offset}) (see \cite[Fig.~3]{Scholl.2026}), immediately allows to conclude the required bound  (\ref{eq:partNegDef})  from   (\ref{eq:kappa_3_existence}). 
Choosing more specifically $\ell(\zeta) =k_3 \|\zeta\|_2^2$ with  $k_3\geq 0$ simply diminishes in (\ref{eq:perturbation_restriction_offset}) the linear norm bound, with the result given in the following corollary.}  

\begin{cor}
\label{cor:smallGainLKFunctional}
Let $\gamma>0$ be chosen such that a functional  $V$  of the form (\ref{eq:V})  exists that solves (\ref{eq:determining_eq_DfV_linear_norm_bound}). 
\rev{Then,} for any desired $k_3\in [0,\gamma^2]$, 
\begin{align} \label{eq:Dfg_k3}
\forall \phi\in \Omega_{k_3}: \quad D_{(f+g)}^+V(\phi)\leq -k_3 \|\mathcal C \phi\|_2^2 
\end{align}
holds on $\Omega_{k_3}\subseteq C$ being the set of all $\phi\in C$ for which   $a(\zeta)$  with $\zeta=\mathcal C\phi$ satisfies the correspondingly reduced linear norm bound 
\begin{align}\label{eq:linear_norm_bound}
\|a(\zeta)\|_2 \leq \sqrt{\gamma^2-k_3} \| \zeta \|_2.  
\\[-2.25em]\nonumber 
\end{align}
\end{cor}
\begin{proof}
For $\ell(\zeta) =k_3 \|\zeta\|_2^2$ in (\ref{eq:kappa_3_existence}),  the strengthened perturbation restriction (\ref{eq:perturbation_restriction_offset}) is (\ref{eq:linear_norm_bound}), cf.\ (\ref{eq:linearNormBound_w}). 
\end{proof}
\rev{In (\ref{eq:linear_norm_bound}), any norm bound slope $ \tilde \gamma:=\sqrt{\gamma^2-k_3}>0$ that is less than~$\gamma$ corresponds to a positive  coefficient $k_3=\gamma^2-\tilde \gamma^2$  to be valid in~(\ref{eq:Dfg_k3}).
If $B=C_0=C_1=I_n$ from (\ref{eq:unstructured}) is chosen, then  
\begin{align} \label{eq:tildegamma}
\|g(\phi)\|_2
&\leq \tilde \gamma \smash{\left\|\left[\begin{smallmatrix} \phi(-\delay) \\ \phi(0)\end{smallmatrix}\right]\right\|_2} 
\qquad
 \text{with }
 \tilde \gamma   
 <   \gamma  
\end{align}
ensures due to  (\ref{eq:linear_norm_bound}) that a positive coefficient $k_3>0$ in (\ref{eq:Dfg_k3}) exists. 
 The constraint (\ref{eq:tildegamma}) on $g$  is the most direct counterpart to (\ref{eq:completeLK_linearNormBound}). 
Note that it relies on the value of  $\gamma$ that is used in (\ref{eq:determining_eq_DfV_linear_norm_bound}), but the maximum admissible value thereof is still open.}

\section{Lower bound on  $V$}\label{sec:properties}

\rev{Having established condition (\ref{eq:partNegDef}) from the LK theorem in the previous section, it remains to consider condition~(\ref{eq:partPosDef}) in this section.} 

\begin{thm}[Lower bound on $V$] 
\label{theorem:global_quadratic}
\rev{Let $\mathrm{rank}(C_1)=\mathrm{rank}(C_0)=n$ in (\ref{eq:C_def}). 
If  the zero equilibrium of the nominal system (\ref{eq:unperturbed}) is asymptotically stable,  then any solution $V$ of   (\ref{eq:determining_eq_DfV_linear_norm_bound}) with $\gamma>0$ 
 satisfies}  
\begin{align}\label{eq:quadraticLowerBound}
\exists k_{1}>0,\forall \phi \in C: \quad   
 k_{1} \|
\phi(0)\|^2\leq V(\phi).  
\\[-2.25em]\nonumber
\end{align} 
\end{thm}
\begin{proof} 
\rev{Let $x_t$ denote the state of  (\ref{eq:unperturbed}) at time $t\geq 0$. By (\ref{eq:kappa_3_existence}),  
\begin{align}
D_{f}^+V(x_t) \leq  -\ell_0(\mathcal C x_t)\quad \text{ with }\;
 \ell_0(\zeta)
:= w(\zeta,0)= \gamma^2 \|\zeta\|_2^2, 
\end{align}
and, thus, for any $t_1>0$, 
\begin{align}\label{eq:diffVineq}
V(x_{t_1})-V(x_0) \leq -\int_0^{t_1} \ell_0(\mathcal C x_t) \,\mathrm dt. 
\end{align}
Take the  argument in $V(\phi)$   as an initial condition $ x_0=\phi \in C([-\delay,0],\mathbb R^n)$ for the nominal system  (\ref{eq:unperturbed}). 
Due to the assumed asymptotic stability of the zero  solution and continuity of $V$, it holds} 
\begin{align}
V(x_0) &=-\Big(\lim_{t_1\to\infty} 
\underbrace{V(x_{t_1})}_{\hspace{-2cm} \to 0
\hspace{-2cm}}
-V(x_0)\Big)
\nonumber \\
& \stackrel{\hspace{-1cm}(\ref{eq:diffVineq})\hspace{-1cm}}\geq \; \rev{\int_0^\infty \rev{\ell_0(\mathcal C x_t)} \,\mathrm d t } 
\nonumber
\\
& \stackrel{\hspace{-1cm}(\ref{eq:C_def})\hspace{-1cm}}=\;
 \int_0^\infty \!\!\rev{\gamma^2}  \|C_0 x(t)\|_2^2 \,\mathrm d t
+  \int_{0}^\infty \! \!\rev{\gamma^2}  \|C_1 x(t-\delay)\|_2^2\,\mathrm d t. 
\nonumber 
\end{align}
\rev{Using the latter}  
as starting point,  the proof   proceeds analogously to the proof of 
\cite[Lem.~2.10]{Kharitonov.2013} from complete-type LK functionals. \rev{That is, based on $\|x(0)\|_2^2=-\lim_{t_1\to\infty} \int_0^{t_1} \frac {\mathrm d}{\mathrm d t} \big(x^\top\!(t)\,x(t)\big)\,\mathrm dt$,   
\begin{align*}
V(x_0)-k_1\|x_0(0)\|_2^2 
\geq
\int_0^\infty \begin{bmatrix} x(t) \\ x(t-\delay)\end{bmatrix}^\top 
\Big(\gamma^2 \begin{bmatrix} C_0^\top C_0 & 0 \\ 0 & C_1^\top C_1 \end{bmatrix}\;\, &
\\
+ 2k_1 \mathrm{sym} \begin{bmatrix} A_0 & A_1 \\ 0 & 0 \end{bmatrix}
\Big)
 \begin{bmatrix} x(t) \\ x(t-\delay)\end{bmatrix}\,\mathrm d t&
\end{align*}  
relies on a positive definite matrix if $k_1>0$ is sufficiently small.}  
\end{proof} 

\rev{Thus, whether condition (\ref{eq:partPosDef}) applies is as well not affected by $\gamma$, the maximal admissible value of which is still open.} 
 
\section{Existence conditions for $V$} 
\label{sec:PerturbationRestrictions}  

\rev{As a last step, we derive conditions that guarantee the existence of a solution $V$ to  (\ref{eq:determining_eq_DfV_linear_norm_bound}), which finally results in a bound on $\gamma$.}

As an intermediate step, we split the solution $V$ of  (\ref{eq:determining_eq_DfV_linear_norm_bound}) 
into a sum $V=V_0+V_1$. \rev{For $V_0$, existence conditions can be proven more simply, whereas $V_1$, which resembles a well-known term from complete-type functionals~\cite{Kharitonov.2013}, exists unconditionally.} 
	
\rev{To shorten notation, we rewrite the $V$-independent term in  (\ref{eq:determining_eq_DfV_linear_norm_bound}) as}
\begin{align} 
&\rev{\gamma^2 (\mathcal C \phi)^\top  (\mathcal C \phi)}
=
\phi^\top\!(-\delay) \underbrace{\rev{\gamma^2  C_1^\top C_1}}_{=:Q_1} \phi(-\delay)
+
\phi^\top\!(0) \underbrace{\rev{\gamma^2 C_0^\top C_0}}_{=:Q_0} \phi(0) . 
	\nonumber
\\[-1.5em]
\label{eq:Q0Q1}
\end{align}
\begin{lem}[Splitting]\label{lem:splitting}
Any  solution $V$ of (\ref{eq:determining_eq_DfV_linear_norm_bound})  
can be split into
\begin{align}\label{eq:V_V0_V1}
V(\phi)&=V_0(\phi)+V_1(\phi),
\quad V_1(\phi)=\int_{-\delay}^0 \phi^\top (\eta) \,Q_1 \phi(\eta)\,\mathrm d \eta,
\end{align}
where $V_0$ satisfies the modified defining equation 
\begin{align} \label{eq:DfV0}
D_{f}^+ V_0(\phi) 
&=
-\phi^\top\!(0) 
(Q_0 +Q_1)\phi(0)
-
\Vvec_0^\top\!(\phi) B B^\top \Vvec_0(\phi)    
\end{align} 
\rev{with $Q_{0}$ and $Q_1$ being defined in (\ref{eq:Q0Q1}).} 
\end{lem}
\begin{proof}
The defining equation (\ref{eq:determining_eq_DfV_linear_norm_bound}) for  $V(\phi)$  \rev{becomes} at $\phi=x_t$ 
\begin{align} 
D_{f}^+ V(x_t) 
&= 
- x^\top\!(t-\delay) 
Q_1 
x(t-\delay)
-x^\top\!(t) 
Q_0
x(t)
\nonumber 
\\* 
&\qquad 
-
\Vvec^\top\!(x_t)  B  B^\top \! \Vvec(x_t) .
\label{eq:DfV_Q0Q1}
\end{align}
It shall be split into $D_{f}^+ V(x_t) =D_{f}^+ V_0(x_t) +D_{f}^+ V_1(x_t)$. For ${\phi=x_t}$, $V_1$ from (\ref{eq:V_V0_V1}) 
reads  $V_1(x_t)=\int_{t-\delay}^t x^\top (\eta) \,Q_1 x(\eta)\,\mathrm d \eta$ with
\begin{align} \label{eq:DfV1}
D_{f}^+ V_1(x_t) 
&= x^\top\!(t) 
Q_1
x(t)
-
x^\top\!(t-\delay) 
Q_1
x(t-\delay).
\end{align} 
Hence, the remaining unknown  $V_0=V-V_1$ 
has to satisfy 
\begin{align} 
D_{f}^+ V_0(x_t) 
&=
-x^\top\!(t) 
Q_0
x(t)
-x^\top\!(t) 
Q_1 
x(t)
\nonumber\\*
&\qquad
- \big(\Vvec_0(x_t)
+
\Vvec_1(x_t)\big)^\top\!  B  B^\top  \big(\Vvec_0(x_t)
+\Vvec_1(x_t)\big)
\nonumber
\end{align} 
where $v(x_t)=v_0(x_t)+v_1(x_t)$ are the corresponding subfunctionals according to (\ref{eq:Vvec_perturbation}). 
In $V_1$ from (\ref{eq:V_V0_V1}), 
the kernel functions in terms of (\ref{eq:V}) are $P_{\mathrm{xx}}=0$, $P_{\mathrm{xz}}(\eta)\equiv 0$, $P_{\mathrm{zz}}(\xi,\eta)\equiv 0$, $P_{\mathrm{zz,diag}}(\eta)\equiv Q_1$,  and thus (\ref{eq:Vvec_perturbation}) yields  
$v_1(x_t)\equiv 0$. 
\rev{Hence,} 
$D_{f}^+ V_0$ becomes (\ref{eq:DfV0}).
\end{proof}

\rev{The defining equation (\ref{eq:DfV0}) for $V_0$ has the advantage not to involve the pointwise  delayed terms $\phi(-\delay)$ that occurred in~(\ref{eq:determining_eq_DfV_linear_norm_bound}).}  
The  following 
existence theorem for $V$ relies on $G\colon \mathbb C \to \mathbb C^{p\times m}$,  
\begin{align}\label{eq:RFDE_G}
G(s) &=\left[\begin{matrix} C_1 \mathrm e^{-s\delay} \\ C_0 \end{matrix}\right]
 (s I - A_0 -\mathrm e^{-s\delay} A_1)^{-1}B,   
\end{align}
involving both the  nominal system (\ref{eq:unperturbed}) and the   structure~(\ref{eq:g_decomp}). 
\begin{thm}[\rev{Existence of $V$}] \label{thm:existence} 
Let the zero equilibrium of~(\ref{eq:unperturbed}) be asymptotically stable. 
If  $\gamma<\gamma_{\max}$ with 
\begin{align}\label{eq:gamma_max}
\gamma_{\max}:=\frac 1{\max_{\omega\geq 0} \|  G(\mathrm i \omega)\|_2} = \frac 1 {\|G\|_{H_\infty}},
\end{align}
where $G(s)$ is given by (\ref{eq:RFDE_G}), then a solution\footnote{Uniqueness of $V$ is not required as the properties proven in the previous sections apply to any solution. If, for some reason, a unique $V$ is desired, the considerations in Appendix \ref{sec:ProofOfExistence} can be restricted to the  so-called stabilizing solution of the ARE (\ref{eq:ARE_P0_infinite_dim}). The latter exists if and only if $\alpha_3>0$ in Lem.~\ref{lem:existence} \cite[Thm.~3]{Likhtarnikov.1977}, i.e., 
the strict inequality $\gamma<\gamma_{\max}$ is even necessary. 
}    $V$ of (\ref{eq:determining_eq_DfV_linear_norm_bound}) exists that has the structure   (\ref{eq:V}). 
\rev{If $\gamma>\gamma_{\max}$, such a solution cannot exist. Moreover, the  structure (\ref{eq:V}) can be made more precise with  $P_{\mathrm{zz,diag}}(\eta)\equiv Q_1$ (see (\ref{eq:V_V0_V1})) being constant.}  
\end{thm}
\begin{proof} 
\rev{See Appendix~\ref{sec:ProofOfExistence}.}
\end{proof}

\section{An LK-based small-gain robustness result} \label{sec:smallgainlike}
The following   counterpart to the robustness statement in  Thm.~\ref{thm:robustness} 
evolves from the previous sections. 
 Numerical implementations of 
\begin{align*}
\|G\|_{H_\infty\!\!}^{}= \max\limits_{\omega\geq 0}\left\| \left[\begin{smallmatrix} \!C_1\mathrm{e}^{-\mathrm i \omega \delay} \!\\ C_0\end{smallmatrix}\right] ( \mathrm i \omega  I_n -A_0-A_1 \mathrm{e}^{-\mathrm i \omega \delay})^{-1} B\right \|_2 
\end{align*}  
are, e.g., available from \cite{Appeltans.2022}, \cite{Michiels.2010b}, \cite{Scholl.2024b}. 

\begin{thm}[\rev{Small-gain robustness statement}]\label{thm:robustnessRobustLK} 
Consider a decomposition  \rev{(\ref{eq:g_decomp})} 
of the perturbation term in (\ref{eq:perturbed_RFDE}) with $\rev{\mathrm{rank}}(C_0)=\rev{\mathrm{rank}}(C_1)=n$. 
Let the zero equilibrium of the nominal linear system~(\ref{eq:unperturbed}) be asymptotically stable. The zero equilibrium of the perturbed system (\ref{eq:perturbed_RFDE}) is still globally asymptotically stable if for all $\zeta\in \mathbb R^p$ the core perturbation function $a(\zeta)$ from (\ref{eq:g_decomp})   resides within the linear norm bound 
$\|a(\zeta)\|_2 \leq \tilde \gamma \| \zeta \|_2$ with  
slope  
\begin{align}\label{eq:smallgainlike}
\tilde \gamma < \tfrac 1 {\|G\|_{H_\infty}}.
\end{align} 
In particular, a corresponding  LK functional $V$ exists 
 that satisfies the conditions from the classical LK theorem (Thm.~\ref{thm:LK}) and that has the form (\ref{eq:V}), with   $P_{\mathrm{zz,diag}}(\eta)$ in (\ref{eq:V}) being constant. 
\nolinebreak Such a functional $V$ is always     found by solving (\ref{eq:determining_eq_DfV_linear_norm_bound}),  
where  $\gamma$ in (\ref{eq:determining_eq_DfV_linear_norm_bound}) can be chosen in the range $\tilde \gamma<\gamma<  1 /{\|G\|_{H_\infty}}$.   
\end{thm}
\begin{proof}
\rev{All conditions of Thm.~\ref{thm:LK} are met:} Existence of an LK functional of robust type $V$ that solves (\ref{eq:determining_eq_DfV_linear_norm_bound}) with $\gamma<\frac 1 {\|G\|_{H_\infty}}$  is ensured by Thm.~\ref{thm:existence}. \rev{Note that $V$ is continuous and upper bounded by Lem.~\ref{lem:upperBound}.}   
It satisfies the \rev{lower bound condition}  (\ref{eq:partPosDef})  by Thm.~\ref{theorem:global_quadratic}, and \rev{the 
condition (\ref{eq:partNegDef}) on $D_{(f+g)}^+V$} by Cor.~\ref{cor:smallGainLKFunctional}, where $k_3=\gamma^2-\tilde \gamma^2$. By Thm.~\ref{thm:existence}, $P_{\mathrm{zz,diag}}$ is constant. 
\end{proof}

 \begin{rem}[Structure]
The LK functional of robust type    has the structure~(\ref{eq:V}) with the last term relying on the constant function  $P_{\mathrm{zz,diag}}(\eta) \equiv \gamma^2 C_1^\top C_1$ (explicitly known due to  (\ref{eq:V_V0_V1})).   Complete-type LK functionals  also have  the structure~(\ref{eq:V}), but the last term relies on a linear function $P_{\mathrm{zz,diag}}(\eta)=Q_1+(\delay+\eta)Q_2$,  \cite{Kharitonov.2013}. A constant $P_{\mathrm{zz,diag}}(\eta)\equiv Q_1$ is \rev{approached only in the non-complete-type limit} $Q_2\to 0_{n\times n}$ in (\ref{eq:determining_eq_DfV_completeTypeLK}), but then  ${\gamma_{\max}\to 0}$ in  (\ref{eq:completeLK_linearNormBound}). 
\end{rem}

\section{Example}\label{sec:example}

We consider the example system 
that,  in  \cite{Kharitonov.2003}, supplemented  the introduction of complete-type LK functionals.

\begin{exmp}\label{ex:original_example}
With a vanishing  perturbation $g(x_t)\equiv 0_n$, 
\begin{align}\label{eq:example_system} 
\dot x(t)&= \begin{bmatrix} 0 & 1 \\ - 1 & -2 \end{bmatrix}
x(t)
+ 
\begin{bmatrix} 0 & 0 \\ -1 & 1\end{bmatrix}
x(t-\delay) +g(x_t)   
\end{align} 
 can be shown to have an asymptotically 
stable zero equilibrium for any $\delay>0$, cf.~\cite{Scholl.2023}. Henceforth, let $\delay=1$ (\!\!\cite[Exmp.~1]{Kharitonov.2003}).

The upper part of Table~\ref{tab:example} compares  linear norm bounds on 
unstructured   perturbations $g(x_t)$ 
and   on  
perturbations 
\begin{align}\label{eq:exampleStructured}
g(x_t) = \begin{bmatrix} 0 \\ g_2(x_t) \end{bmatrix} 
\end{align}
that only affect the second component. 
\rev{The latter are natural,} given (\ref{eq:example_system})  represents the state space description of a second order system for ${y=x_1}$. The structural information is incorporated by choosing \rev{$B=\begin{bmatrix} 0 & 1 \end{bmatrix}^\top$} in (\ref{eq:g_decomp}), thus affecting via  (\ref{eq:RFDE_G}) the resulting bound on $\tilde \gamma$ from (\ref{eq:smallgainlike}).

The lower part of Table~\ref{tab:example} considers 
uncertainties $\Delta_0,\Delta_1\in \mathbb R^{n\times n}$ in the coefficient matrices of (\ref{eq:example_system}) that amount to  \begin{align}\label{eq:gDelta}
g(x_t)=\Delta_0 x(t) + \Delta_1 x(t-\delay). 
\end{align}
The latter can be  decomposed into    
\begin{align*}
g(x_t)&=
[\tfrac 1 {c_1} \Delta_1 
\;\; \tfrac 1 {c_0} \Delta_0 
]
\left[\begin{smallmatrix} c_1 x(t-\delay) \\ c_0 x(t)\end{smallmatrix}\right] 
=:-a\left(\left[\begin{smallmatrix} c_1 x(t-\delay) \\ c_0 x(t) \end{smallmatrix}\right ]\right) 
\end{align*}
with $c_0,c_1>0$. Therefore, choosing $B=I$ and $C_0=c_0I$, $C_1=c_1I$ in (\ref{eq:g_decomp}),  
the core perturbation function is $a(\zeta)=-[\tfrac 1 {c_1} \Delta_1 
\;\; \tfrac 1 {c_0} \Delta_0 
]\zeta$. 
Since $\|a(\zeta)\|_2
 \leq r(\Delta_0,\Delta_1)\, \|\zeta\|_2$ with 
 \begin{align*}
r(\Delta_0,\Delta_1) :=\sqrt{\tfrac 1 {c_1^2} \|\Delta_1\|_2^2+ \tfrac 1 {c_0^2} \|\Delta_0 \|_2^2} 
, 
\end{align*}
a linear norm bound on $a$ is established if $r(\Delta_0,\Delta_1) \leq \tilde \gamma$. Corresponding bounds on $\tilde \gamma$ are calculated from  (\ref{eq:smallgainlike}). \rev{The plot in Table~\ref{tab:example}~(ii) describes resulting admissible combinations of $\|\Delta_0\|_2$ and~$\|\Delta_1\|_2$ in (\ref{eq:gDelta}).} 
\end{exmp}

\rev{Altogether, already in the unstructured case (i)(a) and even more in the structured cases (i)(b) and (ii), the example shows  significantly improved results compared to the ones derived from  complete-type or related LK functionals. }

\renewcommand{\arraystretch}{1.5}
\begin{table}
\centering
\iftoggle{IsTwocolumn}{ 
	\begin{tabular}{p{0.2cm} p{7.7cm}}
	}{\begin{tabular}{p{0.5cm} p{10cm}}}
\hline
(i)& 
\iftoggle{IsTwocolumn}{ 
	\begin{tabular}{>\centering p{7.7cm}}
		}{\begin{tabular}{>\centering p{10cm}}}
Linear norm bound on admissible perturbations  in (\ref{eq:example_system}),  
\\ 
$\|g(x_t)\|_2\leq \tilde \gamma \left\|\left[\begin{smallmatrix} x(t) \\ x(t-\delay)\end{smallmatrix}\right]\right\|_2$ with  $\tilde \gamma<\gamma_{\max}$
\end{tabular} 
\newline
\\
&
\iftoggle{IsTwocolumn}{ 
	\begin{tabular}{@{}p{5.165cm}|p{2cm}} 
	}{\begin{tabular}{@{}p{7.165cm}|p{2.3cm}}}
\hline
LK functional of complete type, Thm.~\ref{thm:robustness}
& $\gamma_{\max}^{}=0.0227$
\\
\hline
LK functional of robust type, Thm.~\ref{thm:robustnessRobustLK} 
\\  
$(a)$ unstructured, $B=C_0=C_1=I$ & $\gamma_{\max}^{}=0.1059$
\\[0.5em]
$(b)$ 
 case of (\ref{eq:exampleStructured}), 
 $B=\left[\begin{smallmatrix} 
0 \\ 1
\end{smallmatrix}\right], C_0=C_1=I$
&
$
\hspace{-0.1em}\gamma_{\max}^{}\hspace{-1.45em}\strut^{g_1\equiv 0\hspace{-0.44em}}
=0.2462$
\end{tabular}
\newline
\\
\hline
(ii)&
\iftoggle{IsTwocolumn}{ 
	\begin{tabular}{>\centering p{7.4cm}} 
	}{\begin{tabular}{>\centering p{9.7cm}} }
Special case of bounds on 
 uncertainties $\Delta_{0,1} \in  \mathbb R^{n\times n}$
 in (\ref{eq:gDelta})    \\ 
yielding   
$\dot x(t)= (A_0+\Delta_0) x(t)+ (A_1+\Delta_1) x(t-\delay)$ in   (\ref{eq:example_system}) 
\end{tabular}
~\newline\vspace{-2em}
\vspace{0pt}   \begin{tikzpicture}
\begin{scope}[scale=15]

\node[anchor=south west,align=left,inner sep=1pt,scale=0.75,
text=blue]   at(0,0.18) {}; 

\begin{scope} 
\path[clip] (0,0) rectangle (0.155,0.155);

\draw[fill=blue!20!white,draw=blue,fill opacity=0.5] (0,0) ellipse [x radius= 0.1490cm, y radius= 0.0149cm]; 

\draw[fill=blue!20!white,draw=blue,fill opacity=0.5] (0,0) ellipse [x radius= 0.0149cm, y radius= 0.1490cm]; 

\draw[fill=blue!20!white,draw=blue,fill opacity=0.5] (0,0) ellipse [x radius= 0.1339cm, y radius= 0.0670cm]; 

\draw[fill=blue!20!white,draw=blue,fill opacity=0.5] (0,0) ellipse [x radius=  0.0670cm, y radius= 0.1339cm]; 

\draw[fill=blue!15!white,draw=blue,fill opacity=0.7] (0,0) circle [radius= 0.1059cm];

\draw[fill=black!50!yellow,draw=black!50!yellow,fill opacity=0.3] (-1,-1) -- ( 0.0481,0) -- (0, 0.0481)  ;  

\draw[fill=green!30!white,draw=green!10!black,fill opacity=0.5] (0,0) circle [radius= 0.0227cm];

\draw[fill=green!50!black,draw=green!10!black,fill opacity=0.4] (-1,-1) -- (  0.0114,0) -- ( 0.0114,0.0122) -- (0,  0.0122)  ;

\end{scope}

\draw[->,thin] (0,0) -- (0,0.17); 
\draw[->,thin] (0,0) -- (0.17,0);

\iftoggle{IsTwocolumn}{ 
	\newcommand\xLeft{-0.29}
	}{\newcommand\xLeft{-0.35}}

\node[anchor=south west,align=left,inner sep=1pt,scale=0.85,  
text=blue]   at(\xLeft,0.08) { 
Thm.~\ref{thm:robustnessRobustLK}, LK f.\ of robust type 
\\
($B=I,\; C_0=c_0 I,\; C_1=c_1 I$) 
};

\draw[-,draw=blue,fill=blue] (\xLeft,0.08) -- (0.025,0.08) circle [radius=0.0015];

\node[anchor=south west,align=left,inner sep=1pt,scale=0.85,  
text=black!70!yellow]   at(\xLeft,0.043) {
\cite[Exmp.~15]{Medvedeva.2015b}, approach from \cite{Medvedeva.2015}  
};

\draw[-,draw=black!50!yellow] (\xLeft,0.043) -- (-0.01,0.043) 
-- (0.005,0.03);
\draw[-,draw=black!50!yellow,fill=black!50!yellow]  (0.005,0.03) circle [radius=0.0015];

\node[anchor=south west,align=left,inner sep=1pt,scale=0.85,  
text=green!70!white!50!black]   at(\xLeft,0.0165) {
Thm.~\ref{thm:robustness}, LK f.\ of complete type  
};

\draw[-,draw=green!70!white!50!black,fill=green!70!white!50!black] (\xLeft,0.0165) -- (0.005,0.0165) circle [radius=0.0015];

\node[anchor=south west,align=right,inner sep=1pt,scale=0.85, 
text=black!70!green]   at(\xLeft,-0.01) {
\cite[Exmp.~1]{Kharitonov.2003},  LK f.\ of complete type 
};

\draw[-,draw=black!70!green] (\xLeft,-0.01) -- (-0.01,-0.01) -- (0.005,0.005);
\draw[-,draw=black!70!green,fill=black!70!green] (0.005,0.005) circle [radius=0.0015];

\draw[stealth-,draw=blue!50!white,text=blue] (0.035,0.115) -- (0.045,0.125)  node[draw=blue!50!white, anchor=south west,align=left,inner sep=2pt,scale=0.65,fill=white,drop shadow,callout relative pointer={(0,1)}]     {
$\begin{aligned}   c_1&=1 \\ c_0&=0.5 \end{aligned}$
};

\draw[stealth-,draw=blue!50!white,text=blue] (0.075,0.075) -- (0.085,0.085)  node[draw=blue!50!white, anchor=south west,align=left,inner sep=2pt,scale=0.65,fill=white,drop shadow,callout relative pointer={(0,1)}]     {
$  c_1=c_0=1$
};

\draw[stealth-,draw=blue!50!white,text=blue] (0.14,0.005) -- (0.15,0.015)  node[draw=blue!50!white, anchor=south west,align=left,inner sep=2pt,scale=0.65,fill=white,drop shadow,callout relative pointer={(0,1)}]     {
$\begin{aligned}   c_1&=0.1 \\ c_0&=1 \end{aligned}$
};

\foreach \x in {0,0.05,0.1,0.15}
\draw[text=black!75!white] (\x cm,0.007cm) -- (\x cm,-0.007cm) node[anchor=north,scale=0.75] {$\x$};

\node[anchor=north,scale=0.75] at (0.2,0.007) {$\|\Delta_0\|_2^{}$};

\foreach \y in {0,0.1}
\draw[text=black!75!white] (0.007cm,\y cm) -- (-0.007cm,\y cm) 
node[anchor=east,scale=0.75] {$\y$};

\node[anchor=east,scale=0.75] at (-0.005,0.14) {$\|\Delta_1\|_2^{}$};

\end{scope}

\end{tikzpicture}
\iftoggle{IsTwocolumn}{ 
	\vspace{0.5em}
	}{\vspace{1.5em}}
\\
\hline
\end{tabular}
\caption{Example~\ref{ex:original_example}. Bounds on admissible perturbations obtained via LK functionals of robust type and via complete-type 
(using $Q_{0,1,2}$ from \cite[Exmp.~1]{Kharitonov.2003}) or related   LK functionals. 
}
\label{tab:example}
\end{table}

\renewcommand{\arraystretch}{1.0}

\section{Conclusion}\label{sec:Conclusion}

The proposed concept \rev{retains the principal} advantage of LK functionals of complete type: an LK functional of robust type that satisfies the conditions of the classical LK theorem is guaranteed to exist whenever the nominal linear system has an asymptotically stable equilibrium. 
At the same time, \rev{the presented} example demonstrates that significantly less restrictive bounds on the norm of admissible perturbations of the time-delay system can be achieved, \rev{matching bounds associated with} small-gain results and the complex stability radius. 
\rev{Future work may build on the presented foundation to derive regional stability statements, which would not be possible from pure frequency-domain arguments. Moreover, the concept can be extended to more general sector-type perturbation constraints. In this respect, the result also admits a QSR-dissipativity interpretation, opening another promising direction for future research.} 

\newcounter{newcounter} %um korrekte Nummerierung von Lemmas etc. im Anhang zu bekommen, startend mit Lemma 11.1 
\setcounter{newcounter}{\arabic{section}}

\iftoggle{IsTwocolumn}{ 
}{\newpage}

\appendix 
 
\setcounter{section}{\value{newcounter}}
\stepcounter{section}
\subsection{Proof of Theorem~\ref{thm:existence}}\label{sec:ProofOfExistence}
\stepcounter{section}

The desired conditions for the existence of an LK functional of robust type $V$ simplify to conditions for the existence of $V_0$  from (\ref{eq:DfV0}) since $V_1$ from (\ref{eq:V_V0_V1}) in   $V=V_0+V_1$ 
 always exists. 
The defining equation (\ref{eq:DfV0}) of $V_0$ 
 can be related to an operator-valued algebraic Riccati equation
 (ARE) in the Hilbert space $L_2([-\delay,0],\mathbb C^n)\times \mathbb C^n$. 
Introducing this ARE is the subject of Sec.~\ref{sec:ProofOfExistence}\ref{sec:operator}. In Sec.~\ref{sec:ProofOfExistence}\ref{sec:Solvability}, solvability of the ARE, and thus existence of $V_0$,  is analyzed via 
 a Hilbert-space version of 
the Kalman--Yakubovich--Popov (KYP) lemma. 
Sec.~\ref{sec:ProofOfExistence}\ref{sec:proofWithG} completes the proof  
of the existence criterion presented in  Thm.~\ref{thm:existence} by providing the link to the transfer function~(\ref{eq:RFDE_G}). 

\subsubsection{Operator-valued ARE in place of the defining equation} 
\label{sec:operator}
We use an embedding into the complex\footnote{\rev{The complexification is needed due to the KYP lemma in Sec.~\ref{sec:ProofOfExistence}\ref{sec:Solvability}.} }  Hilbert space 
 \begin{align} 
M_2 =  L_2([-\delay,0],\mathbb C^n) \times \mathbb C^n
\end{align}
by considering, for any $\phi\in C$, the pair 
\begin{align*}
\left[\begin{matrix} \phi \\ \phi(0) \end{matrix} \right] \in C([-\delay,0],\mathbb R^n) \times \mathbb R^n \subset 
M_2 
\end{align*}
as an element of $M_2$. 
The  inner product\footnote{
We follow the convention to define 
inner products conjugate linear in the second argument. For instance, $\langle r_1,r_2\rangle_{\mathbb C^n}=r_1^\top \overline r_2=r_2^H r_1$.
} in $M_2$ reads
\begin{align}\label{eq:M2innerProduct}
\left\langle \left[\begin{matrix} \phi_1 \\ r_1 \end{matrix} \right] 
,
\left[\begin{matrix} \phi_2 \\ r_2 \end{matrix} \right] 
\right \rangle_{M_2}
= \int_{-\delay}^0 \!  (\phi_2(\theta))^H \phi_1(\theta)  \,\mathrm d \theta
+r_2^H r_1, 
\end{align} 
$\phi_1,\phi_2\in L_2$, $r_1,r_2\in \mathbb C^n$. 
\begin{lem}[Quadratic form in $M_2$] \label{lem:quadraticForm}
Let $P_{\mathrm{xx}}\in \mathbb R^{n\times n}$;  $P_{\mathrm{xz}}\in L_2([-\delay,0],\mathbb R^{n\times n})$; $P_{\mathrm{zz}}\in L_2([-\delay,0]\times [-\delay,0] ,\mathbb R^{n\times n})$ with $P_{\mathrm{xx}}=P_{\mathrm{xx}}^\top$ and $P_{\mathrm{zz}}(\xi,\eta)=P_{\mathrm{zz}}^\top(\eta,\xi)$. Any   quadratic functional of the form  
\begin{align}\label{eq:V0}
V_0(\phi) 
& =
\phi^H\!(0) \,P_{\mathrm{xx}} \,\phi(0)
+
2 \mathrm{Re}\int_{-\delay}^0 \phi^H\!(0)\, P_{\mathrm{xz}}(\eta) \hspace{0.5pt}\phi(\eta)\,\mathrm d \eta
\iftoggle{IsTwocolumn}{ 
	\nonumber \\*
	&\quad+
	}{+}
\int_{-\delay}^0\int_{-\delay}^0 \phi^H\!(\xi) P_{\mathrm{zz}}(\xi,\eta) \hspace{0.5pt}\phi(\eta)\,\mathrm d \eta\,\mathrm d\xi
\end{align}
can be written as a quadratic form in $M_2$
\begin{align}\label{eq:V0_quadForm_M2}
V_0(\phi) = \left\langle \mathscr P_0 \left[\begin{matrix} \phi \\ \phi(0) \end{matrix} \right], \left[\begin{matrix} \phi \\ \phi(0) \end{matrix} \right] \right \rangle_{M_2} 
\end{align}
with a real bounded self-adjoint operator $\smash{\mathscr P_0\colon M_2\to M_2}$,  
\begin{align}\label{eq:operator_P}
\mathscr P_0 
\begin{bmatrix} \phi \\ r  \end{bmatrix}  
&=
\begin{bmatrix}
\mathscr P_{zz}^{}
\phi + \mathscr P_{zx}^{} r 
\\  
\mathscr P_{zx}^* \phi +  
P_{\mathrm{xx}}r 
\end{bmatrix}
= 
\begin{bmatrix}\tilde \phi  \\ \tilde r  \end{bmatrix},
\\*
&\text{with }
\begin{aligned}[t]
\tilde \phi (\theta) 
&=
\smallint_{-\delay}^0 
P_{\mathrm{zz}}(\theta,\eta) \phi(\eta)\,\mathrm d \eta 
+ 
(P_{\mathrm{xz}}(\theta))^H r,
 \\* 
\tilde r &=
\smallint_{\smash{-\delay}}^0 P_{\mathrm{xz}}(\eta) \phi(\eta)\,\mathrm d \eta
+ 
P_{\mathrm{xx}}r   
\end{aligned}
\nonumber
\end{align}
(in terms of 
$\mathscr P_{zz}^{}\!\colon L_2 \to L_2$ and  $\mathscr P_{zx}^{}\colon \mathbb C^n\to L_2$).  
\end{lem}
\begin{proof}
Writing out (\ref{eq:V0_quadForm_M2}) with (\ref{eq:operator_P}) according to  (\ref{eq:M2innerProduct}).
\end{proof}

\begin{rem}[\rev{Structural knowledge}] \label{rem:traceclass}
In particular, the  suboperator $\mathscr P_{zz}\colon L_2\to L_2$  of $\mathscr P_0$ in (\ref{eq:operator_P}) is an integral operator. 
\rev{An operator $\mathscr P_0$ on $M_2$ has the structure (\ref{eq:operator_P}) if and only if it is a Hilbert-Schmidt operator (see, e.g., \cite[Thm.~Vl.23]{Reed.1980}). 
Therefore,  the statement that a Hilbert-Schmidt operator  $\mathscr P_0$  (more specifically even a trace class operator, cf.\ \cite[p.~102/103]{Gibson.1983}) exists, which defines $V_0$ via (\ref{eq:V0_quadForm_M2}),}  immediately implies that $V=V_0+V_1$ from the splitting approach   (Lem.~\ref{lem:splitting}) has the desired structure  (\ref{eq:V}). 
\end{rem}
\begin{rem}[\rev{Role of the splitting approach: structure}] \label{rem:noncompact} Also the overall functional $V=V_0+V_1$  could   be written as a quadratic form 
in $M_2$. However,  
\begin{align*}
V_1(\phi)=\int_{-\delay}^0 \phi^H (\eta) \,Q_1 \phi(\eta)\,\mathrm d \eta = \left\langle \mathscr P_1 \left[\begin{matrix} \phi \\ \phi(0) \end{matrix} \right], \left[\begin{matrix} \phi \\ \phi(0) \end{matrix} \right] \right \rangle_{M_2}
\end{align*}
 from (\ref{eq:V_V0_V1}) relies on a multiplication operator $\mathscr P_1:M_2\to M_2$,  
\begin{align}
\mathscr P_1 
 \left[\begin{matrix} \phi \\ r  \end{matrix} \right] =  
 \left[\begin{matrix} Q_1 \phi \\ 0_n  \end{matrix} \right],
\end{align}
which is not a compact operator in $M_2$. 
\rev{In contrast to Rem.~\ref{rem:traceclass}, the mere statement that a noncompact\footnote{on any infinite-dimensional Hilbert space,   $\{\text{trace class operators}\}\subset \{\text{Hilbert--Schmidt operators}\}\subset \{\text{compact operators}\}.$} operator $\mathscr P=\mathscr P_0+\mathscr P_1$ exists would be  rather vague in terms of the possible structure of its quadratic form $V$.}  
\end{rem}

As a consequence of Lem.~\ref{lem:quadraticForm}, a defining equation for $V_0$ 
should  translate  into a defining equation for~$\mathscr P_0$. 
Note that the left-hand side of the defining equation (\ref{eq:DfV0}), $D_{f}^+ V_0(\phi)$,  is the derivative of $V_0$ along solutions of the nominal linear system $\dot x(t)  =A_0 x(t) + A_1 {x(t-\delay)}$. 
It is well known that such an  RFDE that describes the evolution of $x(t)=x_t(0)\in \mathbb R^n$  can be extended to an abstract differential equation that describes the evolution of $\left[\begin{smallmatrix} x_t \\ x_t(0) \end{smallmatrix} \right]\in M_2$. Namely,
\begin{align}\label{eq:abstractODE}
\tfrac{\mathrm d}{\mathrm d t} \left[\begin{matrix} x_t \\ x_t(0) \end{matrix} \right]= \mathscr A  \left[\begin{matrix} x_t \\ x_t(0) \end{matrix} \right],  \quad t>0,
\end{align}
where 
the operator $\mathscr A \colon D(\mathscr A )\to M_2$ 
\begin{align}\label{eq:infinitesimalGenerator}
\mathscr A   \left[\begin{matrix} \phi \\ r \end{matrix}\right]  
&=
 \left[\begin{matrix} \phi' \\ A_0 r + A_1 \phi(-\delay)
 \end{matrix}\right]  ,
\\*
D(\mathscr A )&= \left\{\left[\begin{smallmatrix} \phi \\ r 
 \end{smallmatrix} \right] \in M_2:
r=\phi(0), \phi'\in L_2, \phi 
\in AC \right \} 
\nonumber \end{align} 
(denoting $\phi'(\theta)=\tfrac{\mathrm d}{\mathrm d \theta} \phi(\theta)$) is the infinitesimal generator of a $C_0$-semigroup on $M_2$. 
See \cite{Curtain.2020} for further details. 

\begin{lem}[Operator-valued ARE] \label{lem:operatorARE}
If $V_0(\phi)$ from (\ref{eq:V0_quadForm_M2}) solves the defining equation (\ref{eq:DfV0}), then 
$\mathscr P_0=\mathscr P_0^*$ solves the ARE 
\begin{align}\label{eq:ARE_P0_infinite_dim}
&\langle \mathscr P_0 \mathscr A  \psi,\psi\rangle_{M_2}
+
\langle   \mathscr A^* \mathscr P_0  \psi,  \psi\rangle_{M_2}
\\
&=
-\langle \mathscr Q  \psi,\psi\rangle_{M_2}
-\langle   \mathscr B^* \mathscr P_0   \psi,    \mathscr B ^* \mathscr P_0 
\psi\rangle_{\mathbb C^m} 
\nonumber 
\end{align} 
$\forall \psi\in D(\mathscr A )$, where $\mathscr A$ is defined in (\ref{eq:infinitesimalGenerator}), and where the real bounded 
operators  
$\mathscr Q\colon M_2\to M_2$, and $\mathscr B\colon  \mathbb C^m\to M_2$ are given by 
\begin{align} \label{eq:mathscrQ}
\mathscr Q \left[\begin{matrix} \phi\\ r\end{matrix} \right]
&=
\left[\begin{matrix} 0_{L_2} \\ (Q_0+Q_1)r
\end{matrix} \right], 
\qquad 
\mathscr B u= \left[\begin{matrix} 0_{L_2} \\ Bu \end{matrix} \right],
\end{align} 
with $\mathscr B ^*  
	\left[\begin{smallmatrix} 
	\phi \\ r 
	\end{smallmatrix} \right] 
	= B^\top r$.
	\end{lem}
	\begin{proof}
The derivative of 
(\ref{eq:V0_quadForm_M2}) along solutions of the abstract differential equation (\ref{eq:abstractODE}) becomes
\begin{align}\label{eq:ARElefthandside}
&D_{f^{\mathrm I}} ^+ V_0(x_t) =D_{\mathscr A} ^+ \left( \langle \mathscr P_0 \! \left[\begin{smallmatrix} x_t \\ x_t(0) 
 \end{smallmatrix} \right],\left[\begin{smallmatrix} x_t \\ x_t(0) 
 \end{smallmatrix} \right]\rangle_{M_2} \right)
 \\*
\iftoggle{IsTwocolumn}{
	&= 
	}{&\hphantom{D_{f^{\mathrm I}} ^+ V_0(x_t)}=}
	\langle \mathscr P_0 \mathscr A \! \left[\begin{smallmatrix} x_t \\ x_t(0) 
 \end{smallmatrix} \right],\left[\begin{smallmatrix} x_t \\ x_t(0) 
 \end{smallmatrix} \right]\rangle_{M_2}
+
 \langle \mathscr P_0\! \left[\begin{smallmatrix} x_t \\ x_t(0) 
 \end{smallmatrix} \right],  \mathscr A \left[\begin{smallmatrix} x_t \\ x_t(0) 
 \end{smallmatrix} \right]\rangle_{M_2} . \nonumber
\end{align}  
The latter with $x_t=\phi$ is the left-hand side of 
(\ref{eq:DfV0}), and the right-hand side 
can analogously be expressed in terms of $\psi= \left[\begin{smallmatrix} \phi \\ \phi(0) 
 \end{smallmatrix} \right]\in M_2$, yielding (\ref{eq:ARE_P0_infinite_dim}).   It is admissible to use the adjoint $\mathscr A^*$   in (\ref{eq:ARE_P0_infinite_dim}) since, relying on the  boundedness\footnote{concerning the role of the boundedness in this respect, cf.\ \cite{Staffans.1996}.} of $\mathscr Q$ and $\mathscr B
^*\mathscr P_0$, it can be shown that  $ \mathscr P_0\psi \in D(\mathscr A^*)$, ${\forall \psi\in D(\mathscr A)}$, see \cite[Cor.~6.5.1]{Curtain.2020}.
	\end{proof}

\begin{rem}[\rev{Role of the splitting approach: KYP applicability}]\label{rem:unboundedQ}
As discussed in Rem.~\ref{rem:noncompact}, the overall functional $V$ could   directly be written in terms of $\mathscr P$. 
Similarly to the above result for $\mathscr P_0$, the operator $\mathscr P$ is  the solution of an ARE. This ARE, however, would involve a term 
\begin{align}
-\phi^H\!(-\delay) Q_1  \phi(-\delay)
&=- \left\langle   \mathscr C_1 \left[\begin{matrix} \phi \\ \phi(0) \end{matrix} \right],  \mathscr C_1 \left[\begin{matrix} \phi \\ \phi(0) \end{matrix} \right] \right \rangle_{\mathbb C^{p_1}}  
\\*
&
 \text{ with }
 \mathscr C_1 \left[\begin{matrix} \phi \\ r \end{matrix} \right]
	=
\rev{\gamma C_1}\phi(-\delay) , 
\label{eq:C1M2}
\end{align} 
\rev{with $\mathscr C_1:M_2\to \mathbb C^{p_1}$ relying on $Q_1=\gamma^2 C_1^\top   C_1$}.  
Due to the pointwise evaluation of $\phi(\theta)$ at $\theta=-\delay$, the operator $\mathscr C_1$ is unbounded in $M_2$. 
For an ARE that involves  an unbounded output operator, neither the structural arguments in Lemma \ref{lem:P0structure} below (cf.\ Rem.~\ref{rem:noncompact}) nor Lemma \ref{lem:infiniteDimKYP}  
 (cf.\ the proof of Prop.~\ref{lem:existence}) in the subsequent section would 
be applicable. 
\end{rem}
 
Conversely to Lem.~\ref{lem:operatorARE}, 
the following lemma ensures that any real bounded self-adjoint  solution  
 $\mathscr P_0$ of (\ref{eq:ARE_P0_infinite_dim}) has the structure given in (\ref{eq:operator_P}). Thus, its quadratic form $V_0$ becomes a functional with the desired structure. 
\begin{lem}[Structure of $\mathscr P_0$]\label{lem:P0structure}
Let  $\mathscr P_0$  be a real bounded self-adjoint operator that solves (\ref{eq:ARE_P0_infinite_dim}). Assume $\mathscr A$ generates an exponentially stable 
 $C_0$-semigroup. 
Then $\mathscr P_0$ has the   
form~(\ref{eq:operator_P}).  
\end{lem}
\begin{proof} 
An analogous result is  known for the stabilizing solution of AREs from standard\footnote{In contrast to standard LQR problems with 
nonnegative costs, the present 
ARE 
would arise in an indefinite LQR problem with input weight 
$R_{\mathrm{LQR}}=I_m$  
but state weight $Q_{\mathrm{LQR}}=-(Q_0+Q_1) 
\preceq 0$.}
 LQR 
problems \cite{Gibson.1983}, and the result can be proven analogously in the present case.  
To this end, note that the right-hand side of 
 (\ref{eq:ARE_P0_infinite_dim}) can be written as $-\langle \mathscr Q_{\mathrm{lyap}} \psi,\psi\rangle_{M_2}\!\!:=\!-\langle \Gamma_1 \psi, \Gamma_1\psi\rangle_{\mathbb C^n}-\langle \Gamma_2\mathscr P_0 \psi, \Gamma_2\mathscr P_0 \psi\rangle_{\mathbb C^m}$  where both 
$\Gamma_1\colon M_2\to \mathbb C^n; 
\Gamma_1 \left[\begin{smallmatrix} \phi\\ r\end{smallmatrix} \right]=
 (Q_0+Q_1)^{\frac 1 2} r$ 
and 
$\Gamma_2\colon M_2\to \mathbb C^m; \Gamma_2 \left[\begin{smallmatrix} \phi\\ r\end{smallmatrix} \right]= B^\top r$ 
are finite rank operators. Therefore,  arguments from \cite[p.~102/103 and Thm.~5.2]{Gibson.1983} apply. In fact, the latter show that $\mathscr P_0$ is a trace class operator \rev{and thereby} establish the structure, see Rem.~\ref{rem:traceclass}.
\end{proof}

As a consequence of the above lemmas,  to obtain a criterion for the existence of an LK functional of robust type $V$, we  only 
need a criterion that ensures solvability of the ARE. More precisely, the conclusion of this section is as follows. 
\begin{prop}\label{prop:AREinsteadOfDef} \rev{Consider $Q_0,Q_1$ from (\ref{eq:Q0Q1}).} Assume that the nominal linear system (\ref{eq:unperturbed})  has an asymptotically 
 stable  equilibrium. 
A solution 
  $\smash{V}$ of  the defining equation (\ref{eq:determining_eq_DfV_linear_norm_bound})  that has the form  (\ref{eq:V})     
	exists if and only if there exists a  real bounded  self-adjoint operator   $\mathscr P_0$ that solves the operator-valued ARE (\ref{eq:ARE_P0_infinite_dim}).  
\end{prop}
\begin{proof}
By Lem.~\ref{lem:operatorARE}, Lem.~\ref{lem:P0structure}, and the splitting approach from Lem.~\ref{lem:splitting}. Note that the stability assumption coincides with the stability assumption on $\mathscr A$ in Lem.~\ref{lem:P0structure} (see also the proof  of Prop.~\ref{lem:existence}).  
\end{proof}

Finally, as a side note, the next lemma shows that the boundedness of the operator $\mathscr P_0$ in $M_2$ implies a quadratic upper bound on  $V$  in~$C$.  
\begin{lem}[Upper bound in $C$]\label{lem:upperBound}
If $V_0$ is described by (\ref{eq:V0_quadForm_M2}) with a bounded operator $\mathscr P_0$ then $V=V_0+V_1$ with $V_1$ from (\ref{eq:V_V0_V1}) satisfies $\exists k_2>0, \forall \phi\in C: V(\phi)\leq k_2 \|\phi\|_C^2$.
\end{lem}
\begin{proof}
In (\ref{eq:V0_quadForm_M2}), $V_0(\phi)\leq 
\|\mathscr P_0\| \big ( \int_{-\delay}^0 \|\phi(\theta)\|_2^2\,\mathrm d \theta + \|\phi(0)\|_2^2\big)
\leq \|\mathscr P_0\| (\delay+1) \|\phi\|_{C,2}^2$, where $\|\phi\|_{C,2}=\max_{\theta\in [-\delay,0]} \|\phi(\theta)\|_2$. 
Moreover, in (\ref{eq:V_V0_V1}), $V_1(\phi)\leq \delay \|Q_1\| \|\phi\|_{C,2}^2$. 
\end{proof}

\subsubsection{Solvability of the operator-valued ARE}
\label{sec:Solvability}
\stepcounter{section}
\stepcounter{subsubsection}\stepcounter{subsubsection} 
The following KYP lemma for $C_0$-semigroups on infinite-dimensional Hilbert spaces is due to  Likhtarnikov and  Yakubovich \cite{Likhtarnikov.1977}. 
\begin{lem}[KYP lemma {\cite[Thm.~3]{Likhtarnikov.1977}}]\label{lem:infiniteDimKYP}
Let $X,U$ be complex Hilbert spaces. Let $\mathscr A\colon X\supseteq D(\mathscr A)\to X$ be the infinitesimal generator of a $C_0$-semigroup, 
and $\mathscr B\colon U\to X$ be a bounded linear operator. Assume $(\mathscr A, \mathscr B)$ is  stabilizable, i.e., there exists a bounded linear operator $\mathscr K_s\colon X\to U$ such that $\mathscr A -\mathscr B \mathscr K_s$ generates an exponentially stable $C_0$-semigroup.  Moreover, assume that  $\mathscr A$ does not have a spectrum in the neighborhood of the imaginary axis. 
Let a  quadratic form   $\mathscr F(x,u)=\langle F_{xx} x,x \rangle_X+ 2\mathrm{Re}\langle F_{ux} x, u\rangle_U +
\langle F_{uu} u,u \rangle_U $ in $X\times U$ be given 
that is continuous. If $\alpha_3>0$, where 
\begin{align} \label{eq:alpha_3}
\alpha_3=\inf_{\omega\in \mathbb R} \inf_{u \in U\setminus\{0_m\}}\frac 1 {\|u\|_U^2} 
\mathscr F\big ((\mathrm i \omega I_{X} - \mathscr A)^{-1} \mathscr B u, u \big), 
\end{align}
then bounded linear operators $\mathscr H=\mathscr H^*\colon  X\to X$ and $\mathscr K\colon X\to U$ exist that satisfy for all $x\in D(\mathscr A)$ and $u\in U$   
\begin{align}\label{eq:Lure_equation_infiniteDim}
2 \mathrm {Re}\langle \mathscr A x+ \mathscr B u, \mathscr H x\rangle_X + \mathscr F(x,u)= \| F_{uu}^{1/2}(\mathscr K x+u)\|_U^2 
\end{align}
(Lur'e equation). If, additionally, $\mathscr A, \mathscr B, \mathscr K_s$ are real, then real operators $\mathscr H,\mathscr K$ exist. 
If ${\alpha_3 <0}$,  such operators cannot exist. 
\end{lem}
Existence of $\mathscr H$ from (\ref{eq:Lure_equation_infiniteDim}) 
establishes existence of $\mathscr P_0$   from (\ref{eq:ARE_P0_infinite_dim})  if $\mathscr F$ is chosen as follows. 
\begin{lem}\label{lem:Lure_ARE}
Let $\psi=x\in X=M_2$, $u\in U=\mathbb C^m$, and  
\begin{align}\label{eq:F_in_infiniteDim_KYP}
\mathscr F  (\psi,u) 
&= 
-\langle \mathscr Q 
\psi,\psi
\rangle_{M_2}
+
\langle   u, u \rangle_{\mathbb C^m}.
\end{align}
Then  the Lur'e equation 
(\ref{eq:Lure_equation_infiniteDim}) with $\mathscr H=-\mathscr P_0$ 
is equivalent to the 
ARE 
(\ref{eq:ARE_P0_infinite_dim}) combined with 
$\mathscr K=- \mathscr B^* \mathscr P_0$. 
\end{lem}
\begin{proof}  With  $\mathscr H=-\mathscr P_0$, $x=\psi$, and with $\mathscr F$ from (\ref{eq:F_in_infiniteDim_KYP}), 
where $F_{uu}=I_m$,  
 equation (\ref{eq:Lure_equation_infiniteDim}) becomes  
\begin{align*}
&-2 \Big(\mathrm{Re}\langle \mathscr P_0  \mathscr A
\psi , \psi\rangle_{M_2}
+
\mathrm{Re}\langle  u , \mathscr B^*\mathscr P_0 \psi \rangle_{\mathbb C^m}
\Big)
\iftoggle{IsTwocolumn}{
	\\*
	&\hspace{3.5cm} -
	}{-}
\langle \mathscr Q \psi, 
\psi
\rangle_{M_2}
+
\langle  u, u \rangle_{\mathbb C^m}
\\
&= 
\langle   \mathscr K \psi ,   \mathscr K \psi  \rangle_{\mathbb C^m}
+ 2 \mathrm{Re}\langle u,    \mathscr K \psi  \rangle_{\mathbb C^m}
 + 
\langle   u ,  u\rangle_{\mathbb C^m}  
.
\end{align*}
Comparing  the mixed terms in $u$ and $\psi$ gives 
$ \mathscr K  = -\mathscr B^*\mathscr P_0 $.
The  terms quadratic in $\psi$ give (\ref{eq:ARE_P0_infinite_dim}). 
\end{proof}

Consequently, Lem.~\ref{lem:infiniteDimKYP} can be used to deduce 
the searched solvability of the operator-valued ARE (\ref{eq:ARE_P0_infinite_dim}). 

\begin{prop}[Existence of $\mathscr P_0$] \label{lem:existence} 
Assume that the zero equilibrium of $\dot x(t)=A_0  x(t)+A_1 x(t-\delay)$  is asymptotically 
 stable. 
A real self-adjoint  bounded operator $\mathscr P_0$ that solves the operator-valued ARE (\ref{eq:ARE_P0_infinite_dim}) exists if $\alpha_3>0$, relying on 
 \begin{align}
\alpha_3&=\inf_{\omega\in \mathbb R} \lambda_{\min}\left(
  W_0(\mathrm i \omega) 
\right),  \quad \text{where} \label{eq:alpha3tildeW}
 \\*
W_0(\mathrm i \omega) &= -(H(\mathrm i \omega))^H (Q_0+Q_1) 
H(\mathrm i \omega)
+I_m,
\label{eq:tildeW}  \\*
H(s) &= (s I - A_0 -\mathrm e^{-s\delay} A_1)^{-1} B. \label{eq:H_I} 
\end{align}
If  $\alpha_3<0$, then such an operator cannot exist. 
\end{prop}
\begin{proof} 
We apply Lemma \ref{lem:infiniteDimKYP} to $\mathscr F$ from (\ref{eq:F_in_infiniteDim_KYP}).  
First, Lemma \ref{lem:infiniteDimKYP} requires $\mathscr F$ to be continuous, which is met 
due to the boundedness of the involved operator $\mathscr Q$ from (\ref{eq:mathscrQ}), 
see, e.g., \cite[Sec.~3]{Kurepa.1987}.  
Second, Lemma \ref{lem:infiniteDimKYP}  imposes a condition  on the spectrum of $\mathscr A$. Note that the assumed asymptotic 
stability implies that the characteristic roots $\{\lambda_k\}_k$ of (\ref{eq:unperturbed}) satisfy $\exists \varepsilon>0, \forall k: \mathrm{Re}(\lambda_k)<-\varepsilon$.  
It is known that these characteristic roots  coincide with the eigenvalues of $\mathscr A$ from (\ref{eq:abstractODE}), which has a mere point spectrum \cite{Curtain.2020}. Moreover, it is known that $\mathscr A$ generates a $C_0$-semigroup that satisfies the spectrum determined growth assumption  
(i.e., ${\sup_k \mathrm{Re}(\lambda_k)<0}$ implies exponential stability of the  $C_0$-semigroup)  \cite{Curtain.2020}. Therefore, not only the condition on the spectrum of $\mathscr A$ 
but also  the stabilizability condition in Lemma \ref{lem:infiniteDimKYP} are satisfied due to the imposed stability assumption. 
The thus applicable existence condition from  Lem.~\ref{lem:infiniteDimKYP} relies on  $\alpha_3$ defined in (\ref{eq:alpha_3}). With $\mathscr A$ given by (\ref{eq:infinitesimalGenerator}), the first argument of $\mathscr F$ in (\ref{eq:alpha_3}) refers to   
\begin{align*}
(s I_{M_2}^{}-\mathscr A)^{-1} \mathscr B
&= 
\left[\begin{matrix} \Phi  \\ \Phi(0) \end{matrix} \right]  
\quad 
\text{ with } \quad
\Phi(\theta) = \mathrm e^{s\theta} H(s),
\end{align*}
 where $H(s)$ is defined in (\ref{eq:H_I}), see \cite[Lem.~7.2.14]{Curtain.2020}. 
Thus, using $\mathscr F$ from (\ref{eq:F_in_infiniteDim_KYP}) and $\mathscr Q$ from (\ref{eq:mathscrQ}), 
\begin{align} 
&\mathscr F  \big((\mathrm i \omega I_{M_2}-\mathscr A
)^{-1} \mathscr B u , u \big)
\label{eq:F_HI} 
\iftoggle{IsTwocolumn}{
	\\* \nonumber 
	&= 
	}{=}
-u^H\!(H(\mathrm i \omega))^H (Q_0+Q_1) 
H(\mathrm i \omega) u
+ u^H  u.  
\end{align}
As a result, $\alpha_3$ from (\ref{eq:alpha_3}) becomes $\alpha_3$ in (\ref{eq:alpha3tildeW}). 
\end{proof}

\subsubsection{Existence criterion of $V$ in terms of $G(s)$}\label{sec:proofWithG}\stepcounter{section}
The existence criterion shall be expressed in terms of $G(s)$ from (\ref{eq:RFDE_G}), rather than $H(s)$ that is used in $\alpha_3$ from (\ref{eq:alpha3tildeW}). \rev{To this end, we rewrite the first term in $ W_0(\mathrm i \omega)$ from (\ref{eq:tildeW}) as follows.} 
\begin{lem}[Dependency on $G(s)$] 
\label{lem:HI_GI_KYP}
\rev{It holds} 
\begin{align}\label{eq:HI_GI_KYP}
(H(\mathrm i \omega))^H (Q_0+Q_1) 
H(\mathrm i \omega)
=
\gamma^2 (G(\mathrm i \omega))^H 
G(\mathrm i \omega)  
\end{align}
\rev{with $H(s)$ from (\ref{eq:H_I}) and $G(s)$ from (\ref{eq:RFDE_G}).} 
\end{lem} 
\begin{proof} 
Starting from the right-hand result in (\ref{eq:HI_GI_KYP}), 
\begin{align*}
\iftoggle{IsTwocolumn}{
	&
}{}
\gamma^2 (G(\mathrm i \omega))^H 
 G(\mathrm i \omega)
\iftoggle{IsTwocolumn}{
	\\
}{}
&=
\gamma^2 (H(\mathrm i \omega))^H 
\Big[
C_1^H \mathrm e^{\mathrm i \omega \delay} \;\; C_0^H  
  \Big]
\begin{bmatrix} C_1  \mathrm e^{-\mathrm i \omega \delay}  \\  C_0   \end{bmatrix} H(\mathrm i \omega) 
\\
&
=(H(\mathrm i \omega))^H (
\gamma^2 C_1^H 
C_1
+\gamma^2 C_0^H 
C_0
) 
H(\mathrm i \omega)
\end{align*}
  makes $Q_0$ and $Q_1$ from (\ref{eq:Q0Q1}) visible. 
\end{proof}

\rev{Altogether, Thm.~\ref{thm:existence} can be proven as follows.}

\begin{proof}\textit{(Proof of Thm.~\ref{thm:existence}.)} 
\rev{By Lem.~\ref{lem:HI_GI_KYP},   $\alpha_3$ from  (\ref{eq:alpha3tildeW}) becomes  
\begin{align*}
\alpha_3&=\inf_{\omega \in \mathbb R} \lambda_{\min}\big(-\gamma^2 (G(\mathrm i \omega))^H 
 G(\mathrm i \omega) + I_m \big) 
\\
&=
-\gamma^2 \sup_{\omega\in \mathbb R} \lambda_{\max}\Big((G(\mathrm i \omega))^H  G(\mathrm i \omega)\Big) +1  
\\
&= -\gamma^2 \|G\|_{H_\infty}^2 +1,
\end{align*}
where $\sup_{\omega\in \mathbb R} \|  G(\mathrm i \omega)\|_2=\|G\|_{H_\infty}$ since $G\in H_\infty$ by the properness of $G$ from (\ref{eq:RFDE_G}) and the assumption on asymptotic stability. 
By Prop.~\ref{prop:AREinsteadOfDef} and Prop.~\ref{lem:existence}, $\alpha_3>0$ (equivalently $\gamma< \gamma_{\max}$ in (\ref{eq:gamma_max}))   implies the searched existence, whereas $\alpha_3<0$ implies nonexistence of $V$. Finally, note that  $\sup_{\omega\in \mathbb R} \|  G(\mathrm i \omega)\|_2=\max_{\omega\geq 0} \|  G(\mathrm i \omega)\|_2$  since $G(\overline s)=\overline{G(s)}$ due to $G$ having real coefficients and since the maximum is attained due to continuity and strict properness of $G$.} 
\end{proof}

\iftoggle{IsTwocolumn}{ 
}{\newpage}

\section*{References}
\bibliographystyle{IEEEtran}
\bibliography{literatur}

\begin{thebibliography}{10}
\providecommand{\url}[1]{#1}
\csname url@samestyle\endcsname
\providecommand{\newblock}{\relax}
\providecommand{\bibinfo}[2]{#2}
\providecommand{\BIBentrySTDinterwordspacing}{\spaceskip=0pt\relax}
\providecommand{\BIBentryALTinterwordstretchfactor}{4}
\providecommand{\BIBentryALTinterwordspacing}{\spaceskip=\fontdimen2\font plus
\BIBentryALTinterwordstretchfactor\fontdimen3\font minus
  \fontdimen4\font\relax}
\providecommand{\BIBforeignlanguage}[2]{{%
\expandafter\ifx\csname l@#1\endcsname\relax
\typeout{** WARNING: IEEEtran.bst: No hyphenation pattern has been}%
\typeout{** loaded for the language `#1'. Using the pattern for}%
\typeout{** the default language instead.}%
\else
\language=\csname l@#1\endcsname
\fi
#2}}
\providecommand{\BIBdecl}{\relax}
\BIBdecl

\bibitem{Boyd.1994}
S.~P. Boyd, \emph{Linear matrix inequalities in system and control
  theory}.\hskip 1em plus 0.5em minus 0.4em\relax Philadelphia: SIAM, 1994,
  vol.~15.

\bibitem{Hale.1993}
J.~K. Hale and S.~M. {Verduyn Lunel}, \emph{Introduction to Functional
  Differential Equations}.\hskip 1em plus 0.5em minus 0.4em\relax New York:
  Springer, 1993.

\bibitem{Kharitonov.2013}
V.~L. Kharitonov, \emph{Time-delay systems: Lyapunov functionals and
  matrices}.\hskip 1em plus 0.5em minus 0.4em\relax New York: {Birkh{\"a}user},
  2013.

\bibitem{Kharitonov.2003}
V.~L. Kharitonov and A.~P. Zhabko, ``{L}yapunov-{K}rasovskii approach to the
  robust stability analysis of time-delay systems,'' \emph{Automatica},
  vol.~39, no.~1, pp. 15--20, 2003.

\bibitem{Medvedeva.2015}
I.~V. Medvedeva and A.~P. Zhabko, ``Synthesis of {R}azumikhin and
  {L}yapunov-{K}rasovskii approaches to stability analysis of time-delay
  systems,'' \emph{Automatica}, vol.~51, pp. 372--377, 2015.

\bibitem{Bajodek.2024}
M.~Bajodek, A.~Seuret, and F.~Gouaisbaut, ``On the necessity of sufficient
  {LMI} conditions for time-delay systems arising from {L}egendre
  approximation,'' \emph{Automatica}, vol. 159, p. 111322, 2024.

\bibitem{Zhabko.2021}
A.~P. Zhabko and I.~V. Alexandrova, ``Complete type functionals for homogeneous
  time delay systems,'' \emph{Automatica}, vol. 125, p. 109456, 2021.

\bibitem{Juarez.2020b}
L.~Ju{\'a}rez, S.~Mondi{\'e}, and V.~L. Kharitonov, ``Dynamic predictor for
  systems with state and input delay: A time-domain robust stability
  analysis,'' \emph{Int. J. Robust Nonlinear Control}, vol.~30, no.~6, pp.
  2204--2218, 2020.

\bibitem{Kudryakov.2023}
D.~A. Kudryakov and I.~V. Alexandrova, ``A new stability criterion and its
  application to robust stability analysis for linear systems with distributed
  delays,'' \emph{Automatica}, vol. 152, p. 110973, 2023.

\bibitem{Juarez.2020}
L.~Ju{\'a}rez, I.~V. Alexandrova, and S.~Mondi{\'e}, ``Robust stability
  analysis for linear systems with distributed delays: A time-domain
  approach,'' \emph{Int. J. Robust Nonlinear Control}, vol.~30, no.~18, pp.
  8299--8312, 2020.

\bibitem{Mondie.2022}
S.~Mondi{\'e}, A.~Egorov, and M.~A. Gomez, ``Lyapunov stability tests for
  linear time-delay systems,'' \emph{Annu. Rev. Control}, vol.~54, pp. 68--80,
  2022.

\bibitem{Alexandrova.2020}
I.~V. Alexandrova, ``On the robustness and estimation of the attraction region
  for a class of nonlinear time delay systems,'' \emph{Appl. Math. Lett.}, vol.
  106, p. 106374, 2020.

\bibitem{Rychkov.2025}
A.~S. Rychkov and A.~Egorov, ``Complete type {L}yapunov-{K}rasovskii
  functionals for the scalar case of general linear delay systems*,'' in
  \emph{IEEE 64th CDC}, 2025, pp. 8340--8345.

\bibitem{MelchorAguilar.2007}
D.~Melchor-Aguilar and S.-I. Niculescu, ``Estimates of the attraction region
  for a class of nonlinear time-delay systems,'' \emph{IMA J. Math. Control
  Inf.}, vol.~24, no.~4, pp. 523--550, 2007.

\bibitem{Medvedeva.2015b}
I.~V. Medvedeva and A.~P. Zhabko, ``A novel approach to robust stability
  analysis of linear time-delay systems,'' \emph{IFAC-PapersOnLine}, vol.~48,
  no.~12, pp. 233--238, 2015.

\bibitem{Villafuerte.2007}
R.~Villafuerte and S.~Mondi{\'e}, ``On improving estimate of the region of
  attraction of a class of nonlinear time delay system,'' \emph{IFAC Proc.
  Vol.}, vol.~40, no.~23, pp. 227--232, 2007.

\bibitem{Popov.1962}
V.~M. Popov and A.~Halanay, ``On the stability of nonlinear automatic control
  systems with lagging argument,'' \emph{Autom. Remote Control}, vol.~23,
  no.~7, p.~7, 1962.

\bibitem{Halanay.1966}
A.~Halanay, \emph{Differential equations: Stability, oscillations, time
  lags}.\hskip 1em plus 0.5em minus 0.4em\relax New York: {Academic Press},
  1966.

\bibitem{Kato.1970b}
J.~Kato, ``Absolute stability of control systems with multiple feedback,'' in
  \emph{Seminar on Differential Equations and Dynamical Systems, Part 2:
  Seminar Lectures at the University of Maryland 1969}, J.~A. Yorke, Ed.\hskip
  1em plus 0.5em minus 0.4em\relax Berlin: Springer, 1970, pp. 128--133.

\bibitem{Walker.1967b}
J.~A. Walker, ``Stability of feedback systems involving time-delays and a
  time-varying non-linearity,'' \emph{Int. J. Control}, vol.~6, no.~4, pp.
  365--372, 1967.

\bibitem{Bliman.2000b}
P.-A. Bliman, ``Extension of {P}opov absolute stability criterion to
  non-autonomous systems with delays,'' \emph{Int. J. Control}, vol.~73,
  no.~15, pp. 1349--1361, 2000.

\bibitem{Bliman.2002c}
------, ``Absolute stability criteria with prescribed decay rate for
  finite-dimensional and delay systems,'' \emph{Automatica}, vol.~38, no.~11,
  pp. 2015--2019, 2002.

\bibitem{Pritchard.1989}
A.~J. Pritchard and S.~Townley, ``Robustness of linear systems,'' \emph{J.
  Differ. Equ.}, vol.~77, no.~2, pp. 254--286, 1989.

\bibitem{Likhtarnikov.1977}
A.~L. Likhtarnikov and V.~A. Yakubovich, ``The frequency theorem for continuous
  one-parameter semigroups,'' \emph{Math. USSR Izv.}, vol.~11, no.~4, p. 849,
  1977.

\bibitem{Likhtarnikov.1977b}
A.~L. Likhtarnikov, ``Absolute stability criteria for nonlinear operator
  equations,'' \emph{Math. USSR Izv.}, vol.~11, no.~5, pp. 1011--1029, 1977.

\bibitem{Likhtarnikov.1976}
A.~L. Likhtarnikov and V.~A. Yakubovich, ``The frequency theorem for equations
  of evolutionary type,'' \emph{Sib. Math. J.}, vol.~17, no.~5, pp. 790--803,
  1976.

\bibitem{Anikushin.2023}
M.~Anikushin, ``Frequency theorem and inertial manifolds for neutral delay
  equations,'' \emph{J. Evol. Equ.}, vol.~23, no.~4, p.~66, 2023.

\bibitem{Scholl.2024c}
T.~H. Scholl, ``Stability in time-delay systems,'' Ph.D. dissertation,
  Karlsruhe Institute of Technology (KIT), Karlsruhe, 2024.

\bibitem{Scholl.2024b}
T.~H. Scholl, V.~Hagenmeyer, and L.~Gröll, ``{L}yapunov-{K}rasovskii
  functionals of robust type and their {L}egendre-tau-based approximation,''
  \emph{IFAC-PapersOnLine}, no.~27, pp. 219--224, 2024.

\bibitem{Khalil.2002}
H.~K. Khalil, \emph{Nonlinear systems}.\hskip 1em plus 0.5em minus 0.4em\relax
  Upper Saddle River: {Prentice Hall}, 2002.

\bibitem{Scholl.2026}
T.~Scholl, ``A {L}yapunov-based perspective on absolute stability,''
  \emph{accepted for at-Automatisierungstechnik}, 2026, arXiv:2606.19311.

\bibitem{Appeltans.2022}
P.~Appeltans, H.~Silm, and W.~Michiels, ``{TDS}-control: a {M}atlab package for
  the analysis and controller-design of time-delay systems,''
  \emph{IFAC-PapersOnLine}, vol.~55, no.~16, pp. 272--277, 2022.

\bibitem{Michiels.2010b}
W.~Michiels and S.~Gumussoy, ``Characterization and computation of {$H_\infty$}
  norms for time-delay systems,'' \emph{SIAM J. Matrix Anal. {\&} Appl.},
  vol.~31, no.~4, pp. 2093--2115, 2010.

\bibitem{Scholl.2023}
T.~H. Scholl and L.~Gr{\"o}ll, ``Stability criteria for time-delay systems from
  an insightful perspective on the characteristic equation,'' \emph{IEEE Trans.
  Automat. Contr.}, vol.~68, no.~4, pp. 2352--2359, 2023.

\bibitem{Reed.1980}
M.~Reed and B.~Simon, \emph{Functional analysis}.\hskip 1em plus 0.5em minus
  0.4em\relax New York: {Acad. Press}, 1980.

\bibitem{Gibson.1983}
J.~S. Gibson, ``Linear-quadratic optimal control of hereditary differential
  systems: Infinite dimensional {R}iccati equations and numerical
  approximations,'' \emph{SIAM J. Control Optim.}, vol.~21, no.~1, pp. 95--139,
  1983.

\bibitem{Curtain.2020}
R.~Curtain and H.~Zwart, \emph{Introduction to Infinite-Dimensional Systems
  Theory}.\hskip 1em plus 0.5em minus 0.4em\relax New York: Springer, 2020.

\bibitem{Staffans.1996}
O.~J. Staffans, ``On the discrete and continuous time infinite-dimensional
  algebraic {R}iccati equations,'' \emph{Syst. Control Lett.}, vol.~29, no.~3,
  pp. 131--138, 1996.

\bibitem{Kurepa.1987}
S.~Kurepa, ``Quadratic and sesquilinear forms. {C}ontributions to
  characterizations of inner product spaces,'' in \emph{Functional Analysis
  II}, S.~Kurepa, H.~Kraljevi{\'c}, and D.~Butkovi{\'c}, Eds.\hskip 1em plus
  0.5em minus 0.4em\relax Berlin: Springer, 1987.

\end{thebibliography}

\end{document}